\documentclass[usenatbib]{mn2e}

\usepackage{color}
\usepackage{times}
\usepackage{graphicx,graphics}
\usepackage{amsmath,amssymb}
\usepackage{natbib}
\title[Oscillating
       shocks in low angular momentum flows]
      {Oscillating
       shocks in the low angular momentum flows as a source of variability of accreting black holes}
\author[P. Sukov\'a and A. Janiuk]
       {P. Sukov\'a\thanks{E-mail: psukova@cft.edu.pl}
        and
       A. Janiuk\thanks{E-mail: agnes@cft.edu.pl}\\
       Center for Theoretical Physics,
       Polish Academy of Sciences,
       Poland}
\begin{document}

\date{}

\pagerange{\pageref{firstpage}--\pageref{lastpage}} \pubyear{}

\maketitle

\label{firstpage}

\begin{abstract}
We derive the conditions for shock formation in a quasi-spherical, slightly rotating flows.
We verify the results of semi-analytical, stationary calculations with the time evolution
studied by numerical hydro-simulations, and we study the oscillations of the shock position. 
We also study the behaviour of flows with varying specific angular momentum, where
the 'hysteresis' type of loop is found when passing through the multiple sonic points region. 
Our results are in agreement with the timescales and shapes of the luminosity flares observed in
 Sgr A*. These models may also be applicable for the Galactic stellar mass black holes,
like GX 339-4 or GRS 1915+105, where periodic oscillations of X-ray luminosity are detected. 

\end{abstract}


\section{Introduction}

Accreting black holes are ubiquitous in the Universe at all scales of masses.
They range from stellar-mass holes found in binary systems, where the accreted material
is transferred from the companion star either via the Roche-lobe overflow through the Lagrange point 
or via the massive, spherical wind, up to the supermassive black holes, which occupy the centers of 
active galaxies. 
The high angular momentum of the accreting gas allows for the formation of a thin disk in the 
plane perpendicular to the rotation axis. If the flow is radiatively efficient, 
the configuration is well described by the classical solution of \cite{1973A&A....24..337S}, 
which is widely used to interpret the optical observations of luminous AGN or soft X-ray spectra of
the black hole binaries, modelled by a multicolor black body emission from the accretion disk. 
Radiatively inefficient flows, on the other hand, 
in which most of the dissipated 
gravitational potential energy is advected onto the central black hole, 
 have been proposed e.g. by \cite{1994ApJ...428L..13N} 
to explain the appearance of low luminosity sources.

Another possibility is that the accretion flow does not possess enough angular 
momentum to form the Keplerian disk, and 
still preserves a spherical, 
or quasi-spherical configuration. {\color{black} In binary systems, such situation could happen if the accretion proceeds through the wind instead of the overflow through the Roche lobe. Also, the sub-Keplerian component could arise from the accretion disc when the advection is important in the inner region. In the case of supermassive black holes, the accretion could be fed by winds from several orbiting stars, which are colliding on the way to the center, thus loosing their angular momentum (for discussion about low angular momentum flows see  \cite{1995ApJ...455..623C} and references therein.)}
Such flows can be at first approximation described by the classical Bondi model \citep{1952MNRAS.112..195B},
in which the radius, given as $R_B=GM/a_\infty^2$, defines 
the so-called sphere of influence of the accreting black hole of mass $M$, when its gravitational potential overcomes
the thermal motions of the intergalactic medium with the sound speed $a_\infty$. Such flows have their luminosities many orders of 
magnitude below the Eddington limit, and may explain the observed flux levels
in the sub-class of Low Luminosity AGN \citep{2003ApJ...582..133D} as well as
in the black hole X-ray  binaries in their quiescent state. 

 \cite{2003ApJ...582..133D} discuss the problem of accretion onto super massive black hole in the galaxy M87 and find that the rate of accretion equal to the Bondi rate matches the overall energetics of the galaxy, based on the estimated jet power.
The tight coupling between the jet power and accretion rate in the nearby elliptical galaxies is however questioned recently by \cite{2011MNRAS.415.3721N} if strong magnetic fields are taken into account. Furthermore, as shown by \cite{2006MNRAS.370L..61P} the magnetic fields can stop the energy release and lead to episodic accretion.
Therefore the accretion rate should be related to the power of magnetically driven jets in a  complex way.

In a physical picture, instead of a pure spherical accretion,
 there will be some amount of angular momentum in the galaxy center. It will be available
for instance from the winds ejected by
 the massive stars that are orbiting the central supermassive back hole, such as observed in
Sgr~A* \citep{2003HEAD....7.0301G,2004MNRAS.350..725L,2008JPhCS.131a2001C}.
{\color{black} However, the amount of angular momentum of the flow is unknown and 
cannot be measured directly, i.e. from dynamical broadening of emission lines. 
The estimated angular momentum is based on the models of winds from emitting 
stars and depends on the assumptions.
Recently, several papers have been published where the low angular momentum model for Sgr~A* accretion was proposed and the amount of angular momentum from wind of the Wolf-Rayet star IRS 13 E3 was estimated to have quite low values $\lambda \in (1.68 \div 2.16)$~$R_{\rm Schw}c$ $\sim (3.36 \div 4.32)M$  \citep{2006MNRAS.370..219M,2012MNRAS.425.2413O}}. {\color{black} However, other 3D simulations show considerably higher values of angular momentum  of stellar wind at the Bondi radius $\sim 10^4M$  \citep{2008MNRAS.383..458C,2014ARA&A..52..529Y}. It is also questionable if the two component advective model possessing both high and low angular momentum segment can apply to Sgr~A* accretion.}

Also, the luminosity of the accreting black holes is not a constant quantity, and the variability on 
various timescales is their common feature.
The low angular momentum flows may also adequately describe the behaviour of Galactic X-ray binaries with
stellar mass black holes, where in their quiescent state the geometrically thin accretion disk
is evaporated or very weak \citep{2014A&A...562A.142M}.
The quasi-spherical flow may also account for the hard energy part 
of the emission of high accretion rate systems, such as GRS 1915+105 \citep{2000ApJ...531L..41C},
where it forms a corona above the thin accretion disk.
For instance, in the power spectrum of Cyg X-2, the low frequency break indicates the need for coronal inflow
\citep{2010LNP...794...17G}. In this picture, further supported i.e. by the lack of variability in the soft black body component in Cyg X-1, 
the red noise component of the observed variability
is defined by viscosity and/or accretion rate fluctuations generated in the coronal flow rather than the 
thin accretion disk.

Hydrodynamical models of the low angular momentum accretion flows have been studied already in 2 and 3 
dimensions, e.g. by
\cite{2003ApJ...582...69P}, \cite{2008ApJ...681...58J} and \cite{2009ApJ...705.1503J}. In those simulations,
a single, constant value of the specific angular momentum was assumed, while the variability of the flows
occurred due to e.g. the non-spherical or non-axisymmetric distribution of the matter. 
The level of this variability 
was also dependent on the adiabatic index.
In these models, the accretion onto black hole proceeds through the poles rather than through the rotationally supported torus, 
so the rate of accretion is a fraction of $\dot{M}_{\rm B}\propto 4 \pi R_{\rm B}^{2}a_{\infty}\rho_{\infty}$, where the dimensionless proportionality constant depends on specific angular momentum value and equation of state.

In the present work, we discuss the hydrodynamical 
models of accretion with a range of specific angular momentum values. In particular, 
we focus on these solutions, where oscillating shocks are found in the gas that passes through the
multiple sonic points, and may manifest themselves in an enhanced emission 
of radiation from that region, observed as flares. 
The shape of the power spectrum characteristic for the source's
 variability as well as the distinct peaks referred
to as quasi-periodic oscillations (QPO) are observables that must be confronted with these 
theoretical models.

Our article is organized as follows. We first describe theoretically the properties of the steady solution of quasi-spherical slightly rotating flow in Section \ref{teorie},  in particular the Rankine-Hugoniot shock condition are studied in Paragraph \ref{teorie_shock}.
Following our analytical results we perform 1D numerical simulations of the accreting flow, which are summarized in Section \ref{results}. The behaviour of the steady shock solutions depending on the parameters is presented in Paragraph \ref{steady}. It turns out that for some part of the  parameter space the shock position is not stable and oscillations of the shock front develop. 
The case, when the angular momentum of the incoming gas from the surroundings is slowly changing due to outer conditions (e.g. when the wind from an orbiting star with eccentric orbit is accreted) is studied in Paragraph \ref{loop}. Our results confirm the existence of the hysteresis loop proposed by \cite{2012NewA...17..254D}.
Finally, we discuss our results in Section \ref{conclusion}, in the frame of the observed properties of accreting black holes in AGN and BHXBs.

\section{Quasi-spherical slightly rotating flow} \label{teorie}
In our study we consider the non-viscous accretion flow with the polytropic equation of state for the gas pressure $p=K\rho^\gamma$, where $\gamma$ is the adiabatic index, hence the local sound speed $a$ is given by the relation 
\begin{equation}
a^2 = \frac{\gamma p}{\rho} = \gamma K \rho^{\gamma-1}, \label{sound_speed}
\end{equation}
where $\rho$ is the gas density. 

Contrary to the recent computations in several papers  (e.g. \cite{2002ApJ...577..880D,2003ApJ...588L..89D,2003MNRAS.343..443D,2006MNRAS.373..146C,2012NewA...17..254D,2012NewA...17..285N}) where thin disc was studied,  we suppose quasi-spherical distribution of the gas, which is endowed by constant specific angular momentum $\lambda$ {\color{black} and which was studied e.g. by \cite{1981ApJ...246..314A}}. Such distribution of matter could be formed instead of an evaporated Keplerian accretion disc and is also suitable for 1D numerical computation. With this difference the computation proceed in similar way like in \citep{2002ApJ...577..880D}, so for the sake of an easy comparison of the results, throughout this section we will consider radial coordinate $r$ to be expressed in the units of the Schwarzschild radius ($R_{\rm Schw} = 2GM/c^2 = 2M$) and angular momentum $\lambda$ in units $R_{\rm Schw}c=2GM/c=2M$, otherwise in other sections as well as in our numerical computations we use geometrical units ($G=c=1,[r]=[t]=[\lambda]=M)$.

For the quasi-spherical flow we obtain the mass accretion rate from the continuity equation in the form
\begin{equation}
\dot{M} = u \rho r^2, \label{continuity}
\end{equation}
where $u$ is the radial inward velocity of the flow.

From (\ref{sound_speed}) density could be written in terms of the sound speed
\begin{equation}
\rho = \left( \frac{a^2}{\gamma K} \right)^n, \quad n=\frac{1}{\gamma-1}. \label{rho_a}
\end{equation}
Thus the mass accretion rate obeys the relation
\begin{equation}
\dot{M} = u r^2 \left( \frac{a^{2}}{\gamma K} \right)^n. \label{mass_rate}
\end{equation}
Accordingly the entropy accretion rate is given by \citep{1989ApJ...347..365C}
\begin{equation}
\dot{\mathcal{M}} = \dot{M} K^n \gamma^n = u \rho r^2 K^n \gamma^n = u r^2 a^{2n}, \label{entropy_rate}
\end{equation}
which is constant everywhere except of the shock position, where entropy is produced and the entropy accretion rate increases discontinuously.

Energy conservation for the steady state could be written in the form
\begin{equation}
\epsilon = \frac{1}{2}u^2+\frac{a^2}{\gamma-1} + \frac{\lambda^2}{2r^2} + \Phi(r), \label{E_cons} 
\end{equation}
where $\Phi(r)$ is the gravitational potential.

Differentiating (\ref{E_cons}) we find that the space gradient of the flow velocity and of the gas sound speed fulfil the equation
\begin{equation}
u\frac{{\rm d}u}{{\rm d}r} + \frac{1}{\gamma-1}2a\frac{{\rm d}a}{{\rm d}r} = \frac{\lambda^2}{r^3} - \frac{{\rm d}\Phi(r)}{{\rm d}r}. \label{du_1}
\end{equation}
The gradient of the sound speed is obtained  from differentiating (\ref{entropy_rate})
\begin{equation}
\frac{{\rm d}a}{{\rm d}r}=(1-\gamma)a\left( \frac{1}{2u} \frac{{\rm d}u}{{\rm d}r} + \frac{1}{r} \right). \label{da_dr}
\end{equation}
Now, substituting (\ref{da_dr}) into (\ref{du_1}) we get the radial gradient of the flow velocity in the form
\begin{equation}
\frac{{\rm d}u}{{\rm d}r} = \frac{\frac{\lambda^2}{r^3}-\frac{{\rm d}\Phi(r)}{{\rm d}r} + \frac{2a^2}{r}}{u-\frac{a^2}{u}} = 
\frac{\frac{\lambda^2}{r^3}- \frac{1}{2(r-1)^2} + \frac{2a^2}{r}}{u-\frac{a^2}{u}}, \label{du_dr}
\end{equation}
where the last equality holds for the Paczynski-Wiita potential, which in the form
\begin{equation}
\Phi(r)=\frac{1}{2(r-1)}
\label{PW_potential}
\end{equation}
mimics the most important features of strong gravitational field near the compact object. The choice of this particular potential is based on \cite{2002ApJ...577..880D}, who claimed that Paczynski-Wiita potential is the best choice of pseudo-Newtonian potential for simulating low angular momentum flows, at least for smaller values of $\gamma$ which are not approaching the nonrelativistic limit $\gamma=5/3$.

We  find the critical points of the flow by making simultaneously the numerator and denominator equal to zero. Hence the flow velocity $u_c$  and sound speed $a_c$ at the critical point satisfy
\begin{align}
u_c&=a_c\label{u_c},\\
a_c&=\sqrt{ \frac{r}{{\color{black}2}} \frac{{\rm d}\Phi(r)}{{\rm d}r} - \frac{\lambda^2}{{\color{black}2}r^2}} = \sqrt{ \frac{r}{{\color{black}4}(r-1)^2} - \frac{\lambda^2}{{\color{black}2}r^2}}\label{a_c}.
\end{align}
Therefore in our case the critical points coincide with sonic points, where subsonic flow with radial Mach number $\mathfrak{M}=u/a < 1$ changes to supersonic flow with $\mathfrak{M} > 1$.
Substituting (\ref{u_c}) and (\ref{a_c}) into (\ref{E_cons}) the equation for the critical point position takes the form
\begin{multline}
\epsilon - \frac{\lambda^2}{2r_c^2}  + \frac{1}{2(r_c-1)}-\\ 
\frac{\gamma+1}{2(\gamma-1)}  \left( \frac{r_c}{{\color{black}4}(r_c-1)^2} - \frac{\lambda^2}{{\color{black}2}r_c^2} \right) = 0. \label{r_c} 
\end{multline}

Solving equation (\ref{r_c}) we find that for some set of parameters ($\gamma,\lambda,\epsilon$) there exist three critical points $r_{\rm in}, r_{\rm mid}, r_{\rm out}$. The middle critical point $r_{\rm mid}$ is of the centre type, the two other are of saddle type, hence the solution of accretion, which flows from infinity down onto the compact center can pass only through $r_{\rm in}$ or $r_{\rm out}$ or both. 

The value of the velocity gradient at the critical point could be obtained as the limit of relation (\ref{du_dr}) for $u \to a$ at $r=r_c$
\begin{equation}
\frac{{\rm d}u}{{\rm d}r}\Big|_{r=r_c} = \sqrt{ \frac{\frac{1}{(r_c-1)^3} + \frac{a_c^2}{r_c^2}({\color{black}4}\gamma - {\color{black}6})- \frac{3\lambda^2}{r_c^4}   } {1 + {\color{black}\gamma}\frac{a_c^2}{u_c^2} } }.\label{du_c}
\end{equation}

\subsection{Shock conditions} \label{teorie_shock}
In case that three critical points exists we have two solutions going through the inner and outer critical point. The solution passing through $r_{\rm out}$ (``outer branch") extends from infinity, where the radial Mach number approach zero, down through $r_{\rm out}$ becoming supersonic there and reaching the center with highly supersonic velocity. The ``inner branch" of solution goes up from the center with supersonic inward speed, passes the inner critical point, which is located quite close to the center, hence it becomes very subsonic in the inner region of accretion. However, because the inner critical point is homoclinic, this solution does not reach infinity, rather it turns around to supersonic motion and returns back to the inner critical point forming a ``fish-shaped" loop (similar to panel (iii) and (iv) in Fig. 2 in \citep{2012NewA...17..254D}). This type of solution can not describe an accretion flow by itself, because the flow is supposed to come from infinity to the center. The accretion flow thus has to follow the outer branch first and then jump to the inner branch of solution  at some point, turning from supersonic motion to subsonic motion abruptly. This phenomenon is called shock. 

Because we assume the flow to be inviscid without any dissipative {\color{black} or radiative} processes {\color{black} (see \cite{1990ApJ...350..281A} for discussion)}, the Rankine-Hugoniot conditions have to be fulfilled at the shock position $r_s$. These conditions express the conservation of mass, energy and momentum in the form
\begin{align}
\dot{M}_+&=\dot{M}_-,\\
\epsilon_+&=\epsilon_-,\label{shock_e}\\
\rho_+u^2_+ + p_+&=\rho_-u^2_-+ p_- \label{shock_moment},
\end{align}
where ``-" subscript belongs to quantities measured before the flow reaches the shock and ``+" subscript stands for the postshock values.

Considering (\ref{E_cons}) equation (\ref{shock_e}) takes the form
\begin{equation}
\frac{\mathfrak{M}_+^2a_+^2}{2} + \frac{a_+^2}{\gamma-1} = \frac{\mathfrak{M}_-^2a_-^2}{2} + \frac{a_-^2}{\gamma-1} ,
\end{equation}
hence the ratio of postshock and preshock sound speed is given by
\begin{equation}
\left(\frac{a_+}{a_-}\right)^2 = \frac{\frac{1}{2}\mathfrak{M}_-^2 + \frac{1}{\gamma-1}}{\frac{1}{2}\mathfrak{M}_+^2 + \frac{1}{\gamma-1}}. \label{a2_I}
\end{equation}

Combining (\ref{rho_a}) and (\ref{entropy_rate}), the density could be expressed in terms of mass and entropy rate
\begin{equation}
\rho=\frac{a^{2n} \dot{M}}{\dot{\mathcal{M}}}. \label{rho_M}
\end{equation}

Substituting $p$ from (\ref{sound_speed}) and $\rho$ from (\ref{rho_M}) into (\ref{shock_moment}) the third shock condition becomes 
\begin{equation}
\frac{a_+^{2n+2}}{\gamma \dot{\mathcal{M}}_+}(\gamma \mathfrak{M}_+^2 + 1) = \frac{a_-^{2n+2}}{\gamma \dot{\mathcal{M}}_-}(\gamma \mathfrak{M}_-^2 + 1) .
\end{equation} 
From (\ref{entropy_rate}) the entropy accretion rate in terms of Mach number and sound speed is given as $\dot{\mathcal{M}}=r^2\mathfrak{M}a^{2n+1}$, which leads to the ratio of the sound speeds in the form
\begin{equation}
\left(\frac{a_+}{a_-}\right)^2 =\left(\frac{\mathfrak{M}_+}{\mathfrak{M}_-}\right)^2 \left( \frac{\gamma \mathfrak{M}_-^2+1}{\gamma \mathfrak{M}_+^2+1}\right)^2 \label{a2_II}.
\end{equation}

Finally, comparing (\ref{a2_I}) and (\ref{a2_II}) we obtain the following shock condition 
\begin{equation}
\frac{ \left(\frac{1}{\mathfrak{M}_+}  + \gamma \mathfrak{M}_+ \right)^2 }{ \mathfrak{M}_+^2(\gamma - 1) + 2 } = \frac{ \left(\frac{1}{\mathfrak{M}_-}  + \gamma \mathfrak{M}_-\right)^2 }{ \mathfrak{M}_-^2(\gamma - 1) + 2 } \label{shock}.
\end{equation}

At the shock position important dynamical quantities change abruptly in very thin shell of gas. The flow becomes subsonic quickly and the increase of density of the accreting matter is described by the compression ratio $R_{\rm comp}=\rho_+/\rho_-$.


\section{Numerical results} \label{results}
\label{results}

\begin{figure*}
\includegraphics[width=0.495\textwidth]{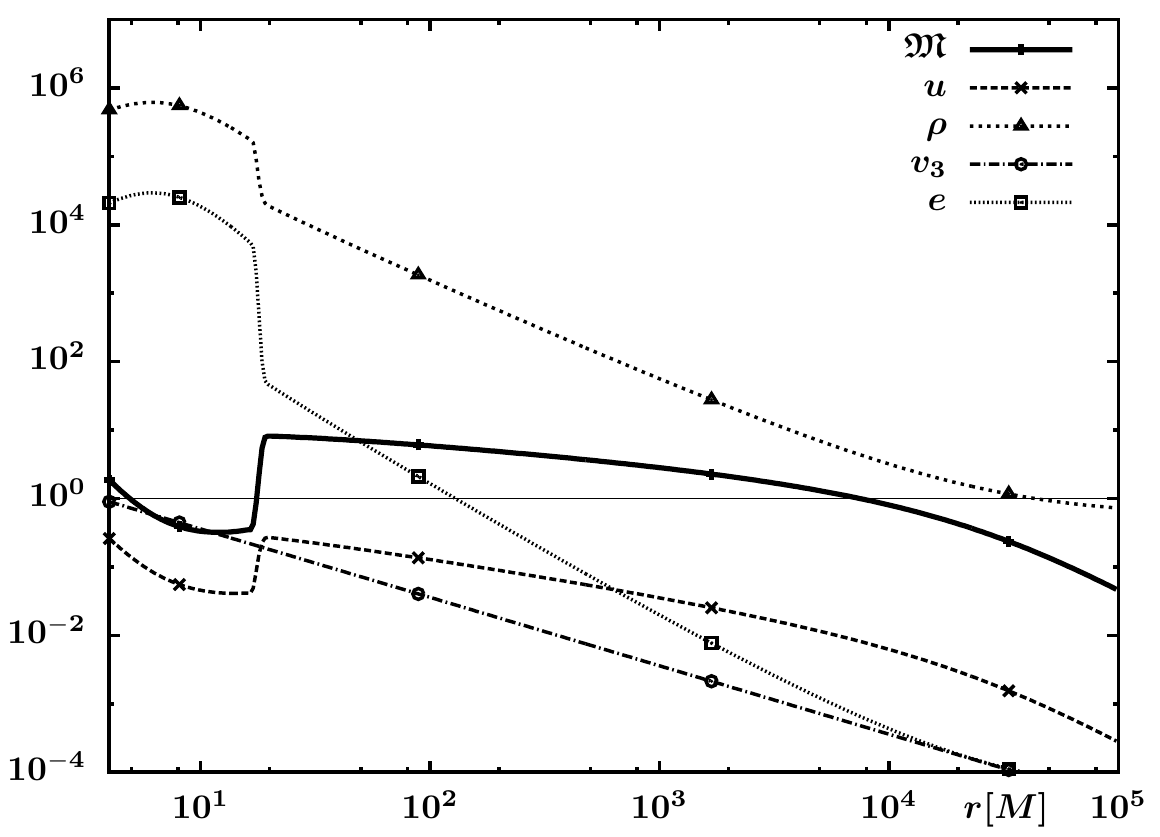}
\includegraphics[width=0.495\textwidth]{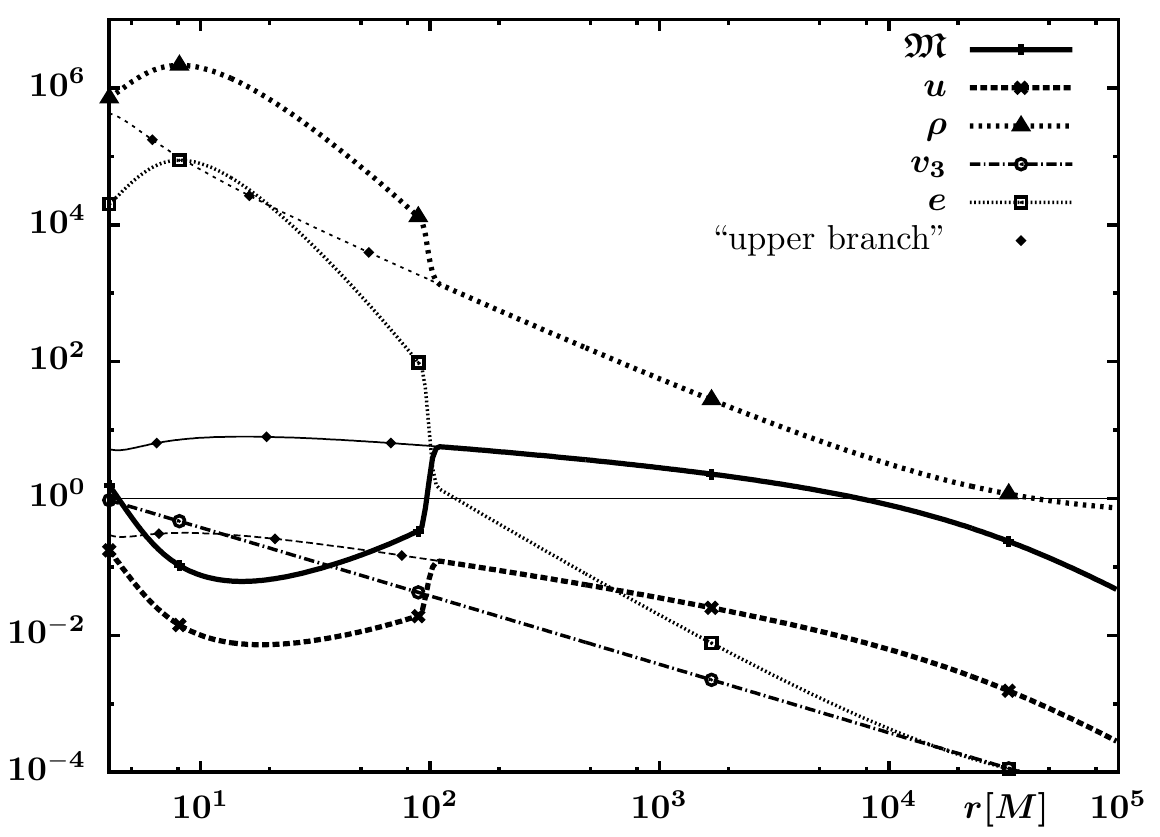}
\caption
{Radial course of the Mach number $\mathfrak{M}$, density $\rho$, inward velocity $u$, angular velocity $v_3$ and internal energy $e$ of the accretion flow solution with standing Rankine-Hugoniot shock.  The parameters of the system are $\gamma=4/3$, $\epsilon=0.0001$, $\lambda=3.6M$ on the left plot and $\lambda=3.78M$ on the right plot. The velocities are plotted in the units of $c$, the density and energy profile is given by the chosen value of accretion rate $\dot{M}=2\,000\,000$ ($\rho_\infty=0.544$). The critical points position is $r_{\rm in}=4.9M, r_{\rm out}=	7485.04M$ in the left plot and  $r_{\rm in}=4.39M, r_{\rm out}=7482.81M$ in the right plot, the shock position is $r_s=20.93M$ in the left plot and  $r_s=110.57M$ in the right plot. The line $y=1$ is plotted for better visibility of the sonic points. {\color{black} In the right plot, the profiles of $\mathcal{M}$, $u$ and $\rho$ of the shock-free outer branch are also depicted (thin lines with diamonds).}
\label{fig_ha_hn}
}
\end{figure*}

For the numerical study of these phenomenas we use the ZEUS-MP code (\cite{1992ApJS...80..753S,2003ApJS..147..197H}), which solves the set of hydrodynamics equations
\begin{align}
\frac{{\rm d}\rho}{{\rm d}t} + \rho \nabla\cdot \vec{v} &=0,\\
\rho\frac{{\rm d}\vec{v}}{{\rm d}t} &=- \nabla p + \rho \nabla \Phi(r),\\
\rho \frac{{\rm d}}{{\rm d}t} \left( \frac{e}{\rho}\right) + p \nabla \cdot \vec{v} &= 0,
\end{align}
where $\vec{v}$ is the velocity of the flow, $e$ is the internal energy, which is related to the pressure by the relation $p=(\gamma-1)e$ and $\text{d}/\text{d}t$ denotes material time derivative. In our case we use the 1D version of code, which means that the grid is spanned only through the interval of radii corresponding to the slice of equatorial plane. During the computation all three components of velocity are considered, the first component equals to the negative value of inward velocity $u=-v_1$, the second component $v_2=0$ and the third component is given by the specific angular momentum $v_3=\lambda / r$.

To take into account the effect of strong gravity near the compact object we use the pseudo-Newtonian gravitational potential (\ref{PW_potential}), which differs significantly from the general relativistic description only at the innermost region around the horizon of the compact star at distances $1r_g $-- $2r_g$. Because the inner boundary of our computation domain is placed at $r_{\rm ib}=4M$ or slightly below this value due to some stability requirement, as will be discuss later, the use of pseudo-Newtonian Paczynski-Wiita potential is sufficient for our purposes.
For treating the inner region of the flow properly we use logarithmic grid, which is denser close to the inner boundary. The outer boundary $r_{\rm ob}$ is placed in the range $\sim(10^4\div 10^5)\,M$, so that it is well beyond the outer critical point for the considered value of energy.

\subsection{Numerical simulations of the shocked flow} \label{steady}
\begin{figure*}
\includegraphics[width=0.495\textwidth]{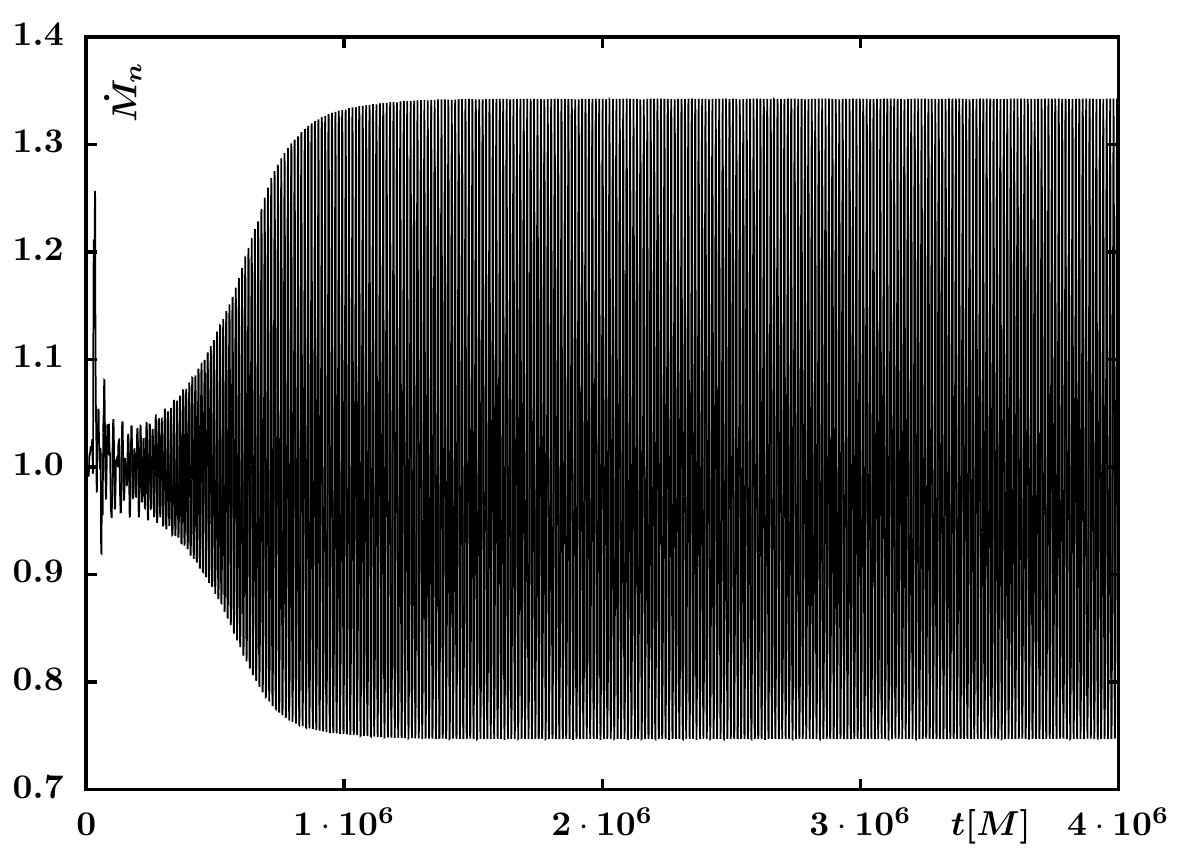}
\includegraphics[width=0.495\textwidth]{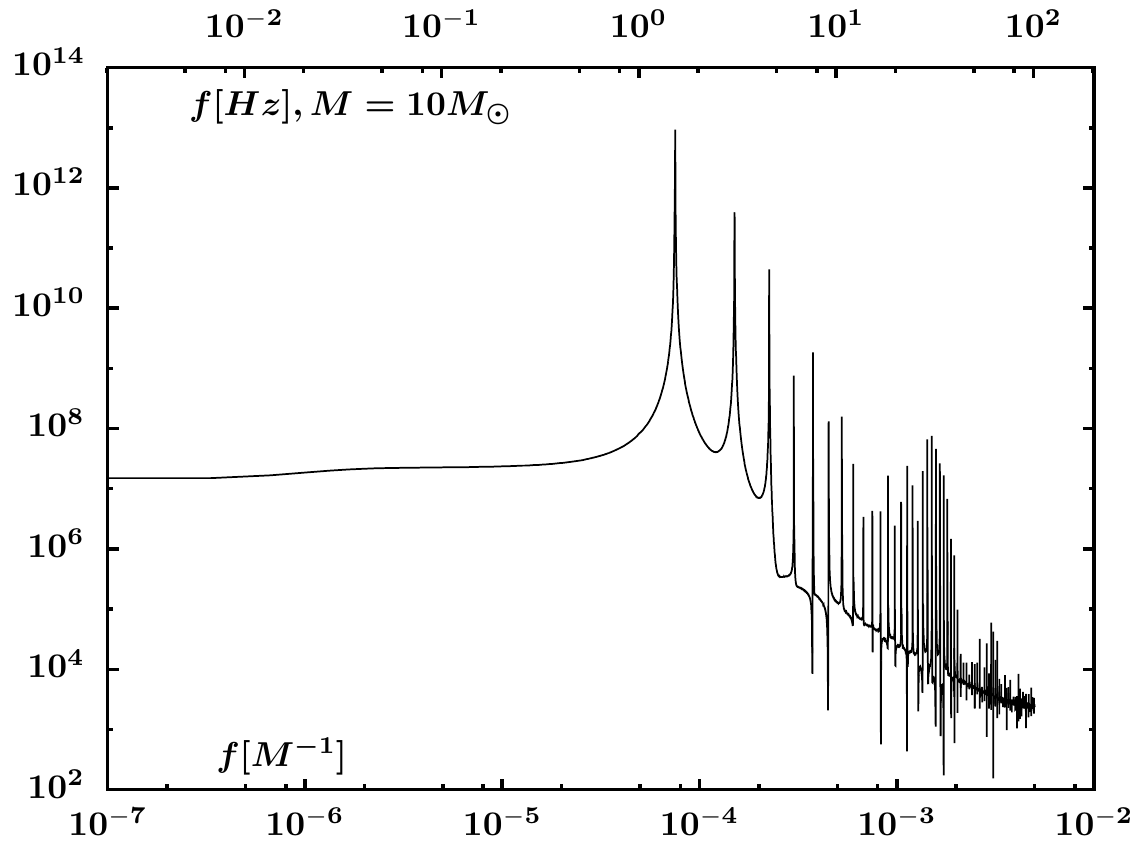}

\includegraphics[width=0.495\textwidth]{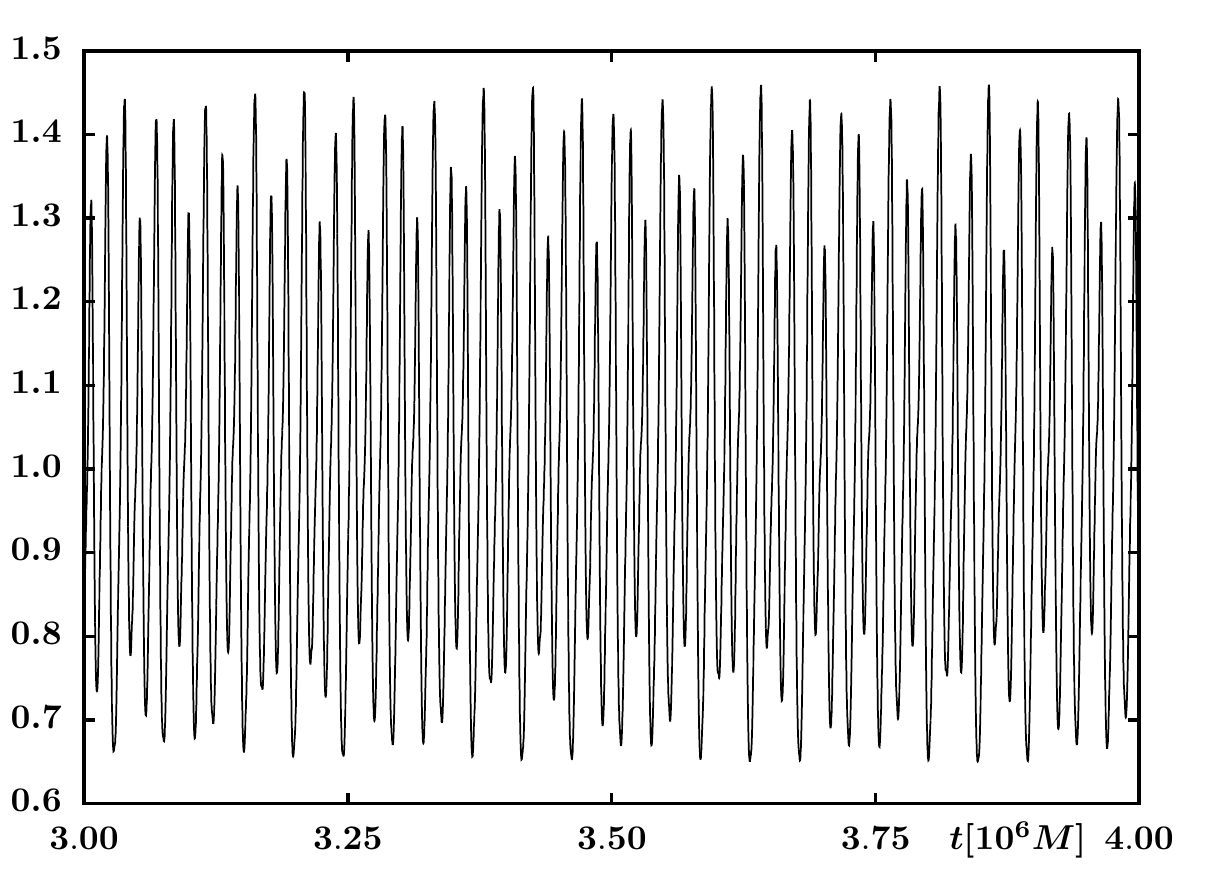}
\includegraphics[width=0.495\textwidth]{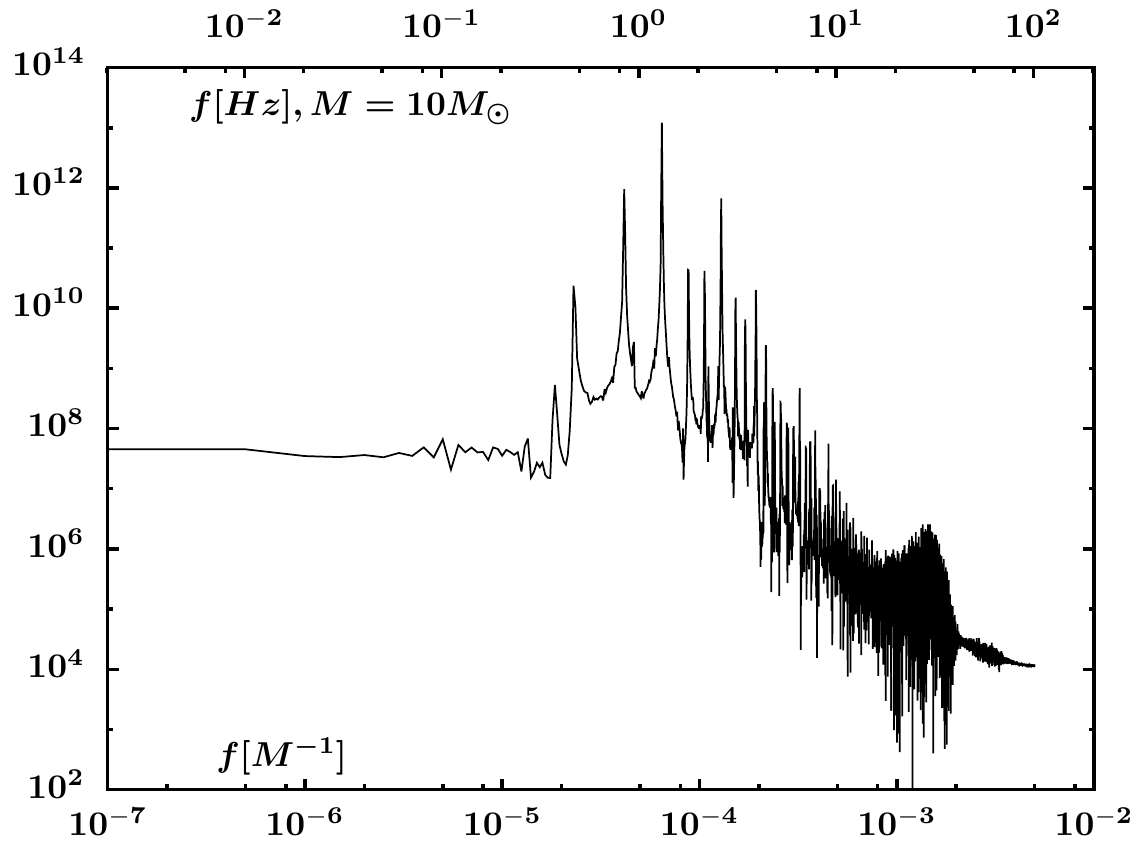}

\includegraphics[width=0.495\textwidth]{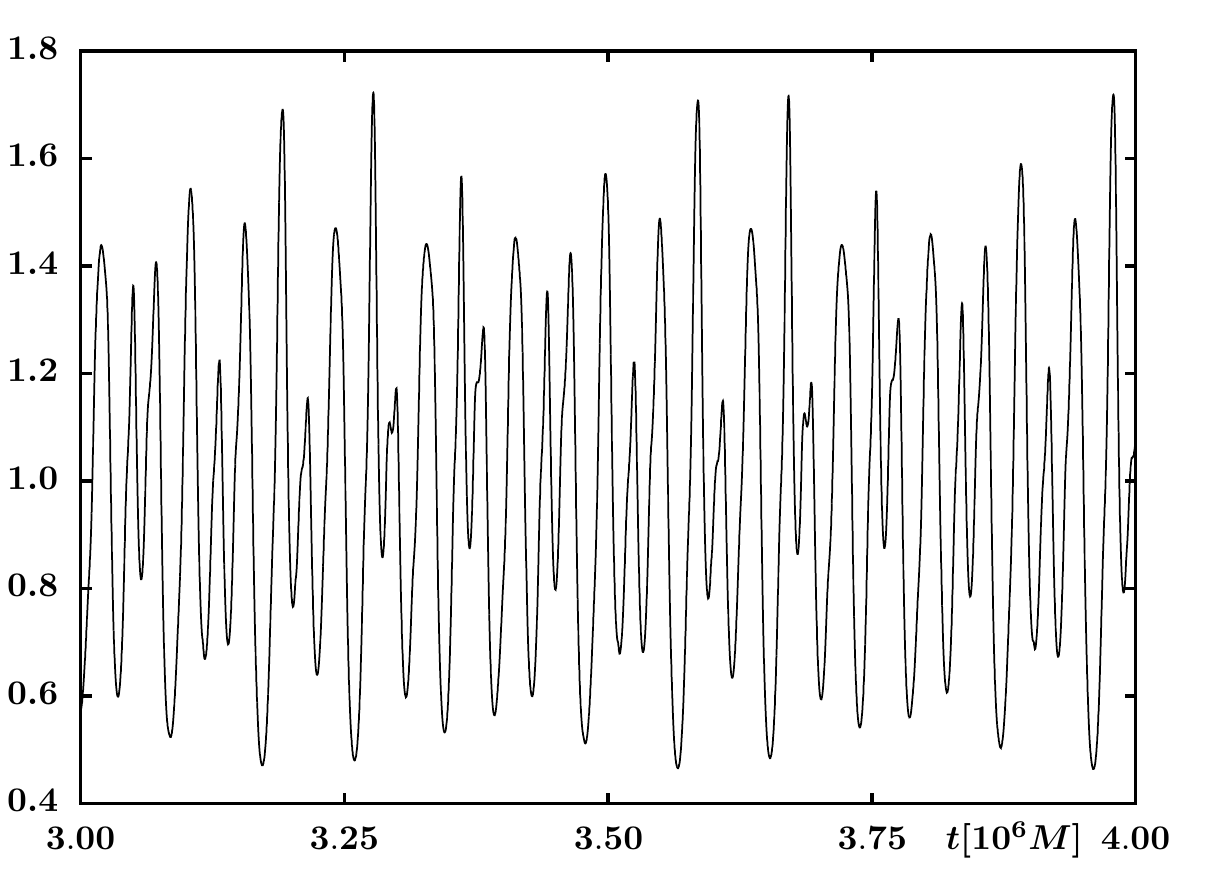}
\includegraphics[width=0.495\textwidth]{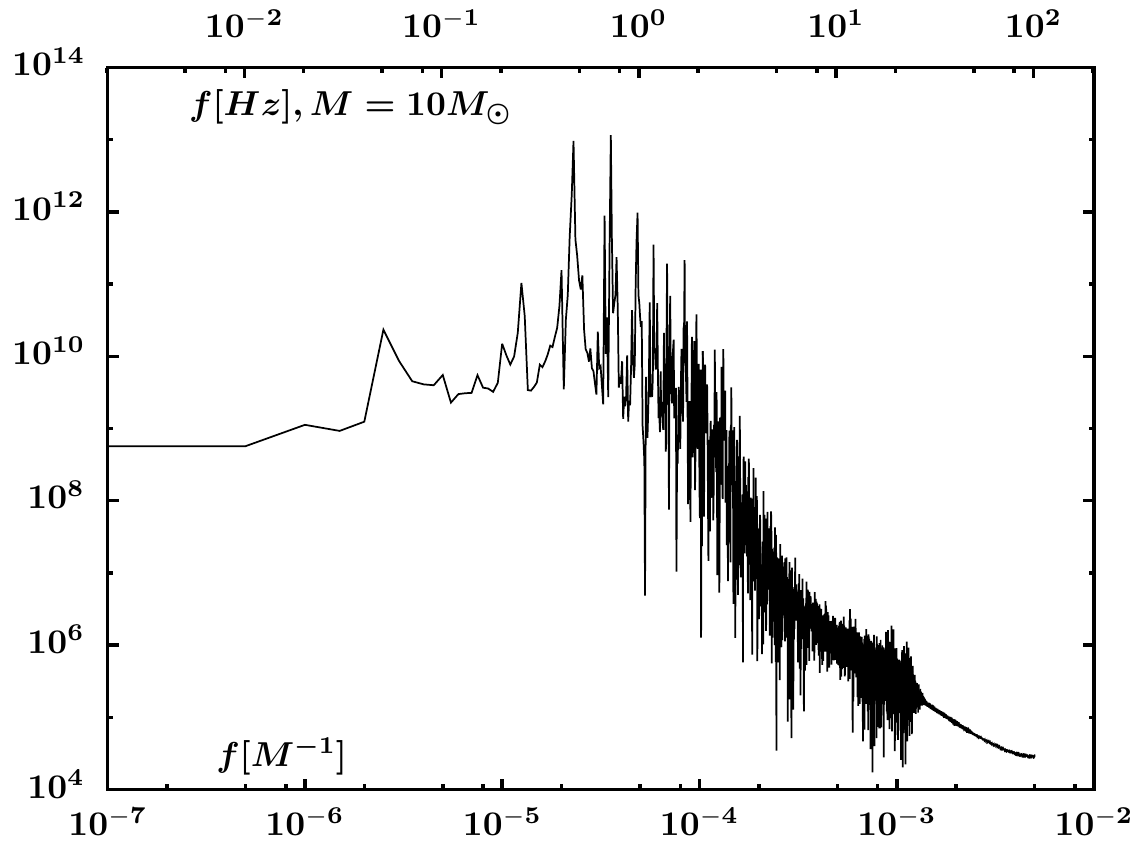}
\caption
{Time variation of normalized accretion rate $\dot{M}_n=\dot{M}(t)/\dot{M}(0)$ through the inner boundary (left column) and the respective power spectrum (right column), which is obtained after the oscillations  stabilize (zoom of the used time period is plotted in the last two rows). The top x axis on the right plots shows the values of frequency in $Hz$ for a representative stellar black hole with the mass $M=10M_\odot$.  The parameters of the system are $\gamma=4/3$, $\epsilon=0.0001$, $\lambda=3.83M$ for the top row, $\lambda=3.835M$ in the middle row and $\lambda=3.855M$ in the bottom row. For the value of specific angular momentum $\lambda=3.855M$ two peaks with similar amplitude could be seen in the spectrum.
\label{fig_ic}
}
\end{figure*}

In this section we confirm and extend the aforesaid semi-analytical results with numerical computation in ZEUS code. We focus on the time evolution of the solutions with standing Rankine-Hugoniot shocks.  {\color{black} In the multicritical region, the accretion can proceed along two possible steady solutions. Either it can go through the outer sonic point only, being supersonic within the inner region of accretion (``outer branch'') or the shock can form and the flow passes through both sonic points (both branches are depicted in the right plot of Fig. \ref{fig_ha_hn}). The choice between these two possibilities depends on the initial conditions of the flow, which have to be prescribed at the beginning of the computation. According to our simulations, the Bondi initial conditions (solution without any angular momentum) with a constant angular momentum gas injected through the outer boundary, leads to the shock-free accretion (outer branch). On the other hand, solution constructed from the inner branch tied to the outer branch at some arbitrary radius (not at the exact shock position) quite quickly converges to the shock solution and the shock position is stable after that. For creation of the shock the conditions in the inner part of the flow seem to be crucial, and the evolution of the flow is sensitive to the existence of the subsonic region near the inner boundary and the presence of the inner sonic point. }

\begin{figure*}
\includegraphics[width=0.495\textwidth]{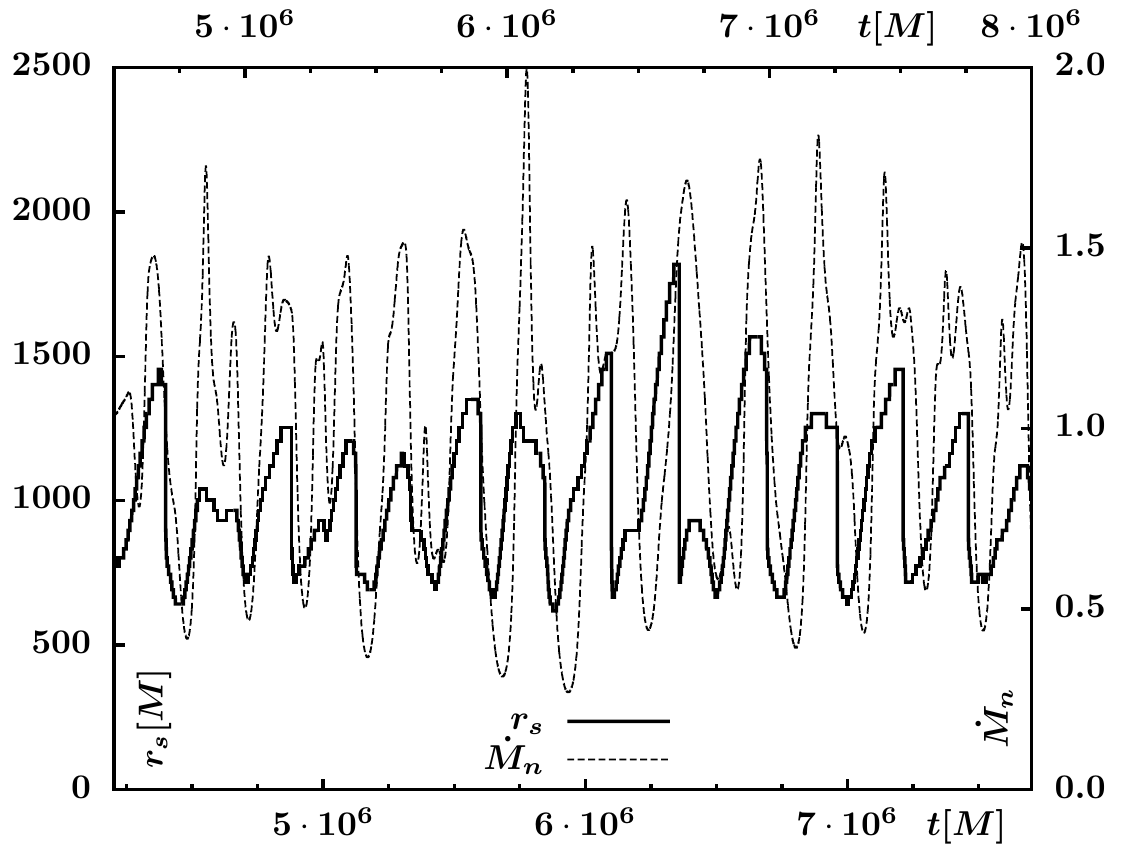}
\includegraphics[width=0.495\textwidth]{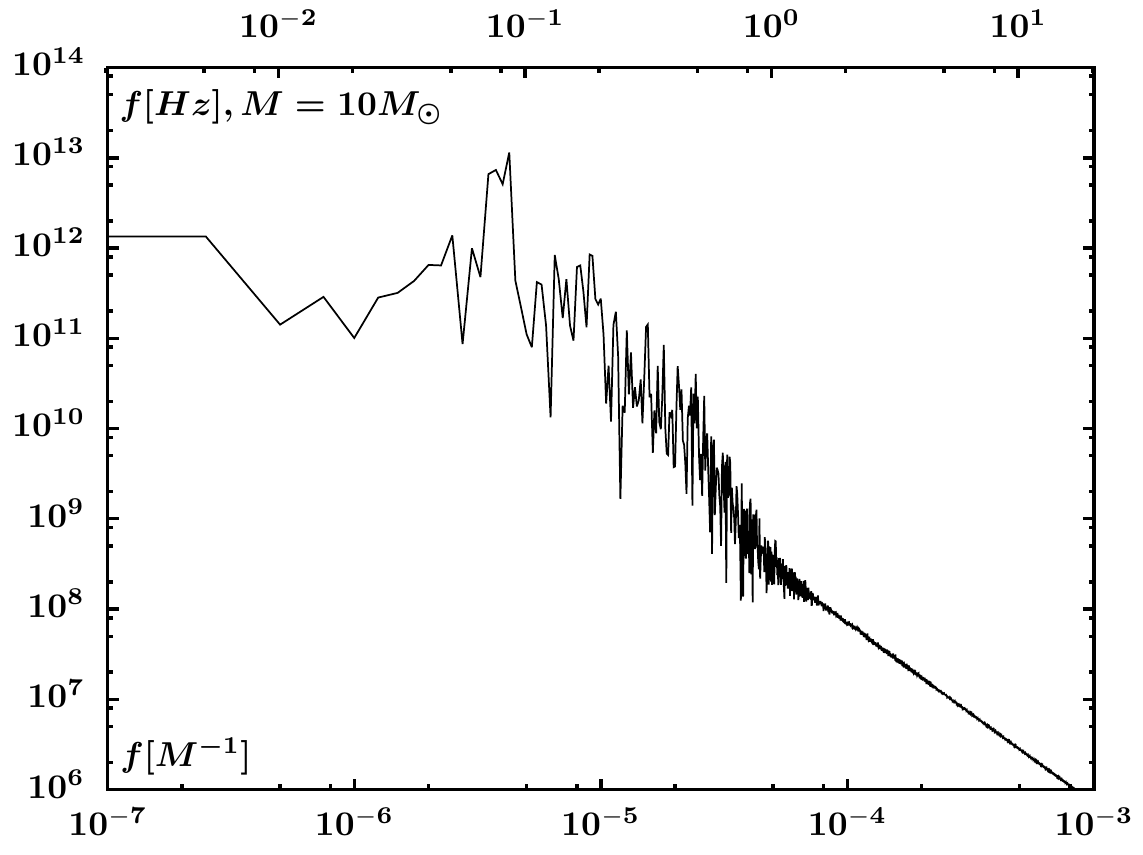}
\caption
{Left: time variation of  the shock position (solid line, bottom x axis) and normalized accretion rate $\dot{M}_n=\dot{M}/\dot{M_0}$ through the inner boundary (dashed line, top x axis). The accretion rate is plotted with the time delay equal to the advection time $\tau_{\rm adv}=293807M$ for the initial position of the shock $r_s=888M$, the time coordinate for the accretion rate is indicated on the top x axis. Right: power spectrum of the accretion rate. Top x axis shows the frequency in $Hz$ for the mass of the compact object $M=10M_\odot$. The parameters of the system are $\gamma=4/3$, $\epsilon=0.0001$, $\lambda=3.9M$. (The animation of this solution evolution is accessible  in the online version.)
\label{fig_hv}
}
\end{figure*}

{\color{black}Because we are interested in the properties of the shock solution, we prescribe it as the initial condition for the flow variables. However, as we will show in Section \ref{loop}, under certain circumstances the shock creates in the shock-free flow during the dynamical evolution, when it is not given by the initial conditions.

 The shock solution is found in the following way. }
For a given set of constant parameters ($\gamma,\lambda,\epsilon,\dot{M}$) we find the inner and outer critical point. According to equations (\ref{u_c}) and (\ref{a_c}) the values of $r_c,u_c$, $a_c$ and $\mathfrak{M}_c$ is known for both critical points. From (\ref{mass_rate}) we determine the two values of the constant $K$ and the entropy accretion rate $\dot{\mathcal{M}}$ for the inner ($K_{\rm in}, \dot{\mathcal{M}_{\rm in}} $) and outer ($K_{\rm out},\dot{\mathcal{M}}_{\rm out}$) branch. Then we numerically compute the outer branch of solution (velocity $u$, sound speed $a$, density $\rho$, internal energy $e$) integrating from the outer critical point to both directions, solving (\ref{du_c}) for $r$ very near to the critical point and (\ref{du_dr}) elsewhere with $a$ given by (\ref{entropy_rate}). We divide the distance between successive zones in the grid to several thousands smaller steps for achieving sufficient accuracy of the integration. By doing so the integration step size follows the logarithmic scale of the grid, so that the step is much smaller near to the inner boundary, where the changes of variables are bigger. When the algorithm arrives at the appropriate mesh position, the corresponding variables in the grid are set to the computed values. 

Moreover, when we integrate from $r_{\rm out}$ downwards to the center, we simultaneously solve the shock condition (\ref{shock}) at every integration step, find the value of $\mathfrak{M}_+$ and the corresponding entropy accretion rate $\dot{\mathcal{M}}_+$ and check if it equals the value of the entropy accretion rate of the inner branch $\dot{\mathcal{M}}_{\rm in}$. Therefore the shock condition is checked on much smaller scale then the size of the zone in the grid.

When the shock condition is met together with the requirement $\dot{\mathcal{M}}_+=\dot{\mathcal{M}}_{\rm in}$, the shock position $r_s$ is found and the computation switches to the inner branch, starting at the inner critical point $r_{\rm in}$ and following the same procedure as above for $r_{\rm ib}<r<r_s$ stopping in the last zone below the shock.

The boundary condition at the inner boundary is set to the outflow boundary condition, which is exact for supersonic outflows, whereas for the subsonic outflows reflection waves occur \citep{1992ApJS...80..753S}. In our case the flow passes through the sonic point and is supersonic near the inner boundary, so the condition is applicable. The outer boundary is set to the inflow boundary condition, hence the values  of the variables in the last two ghost zones are set to constant values computed from the initial procedure described above. Eventhough the inflow is subsonic, no spurious waves of energy occur.

\begin{figure*}
\includegraphics[width=0.495\textwidth]{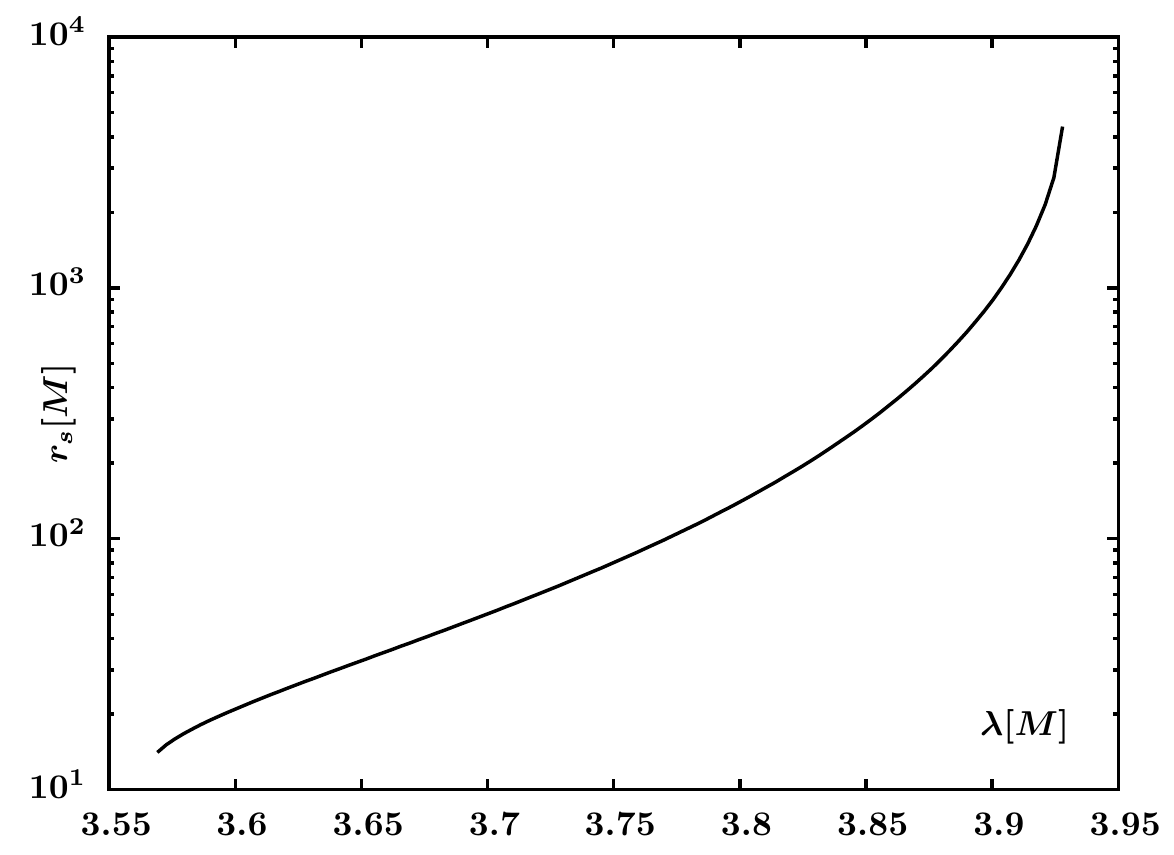}
\includegraphics[width=0.495\textwidth]{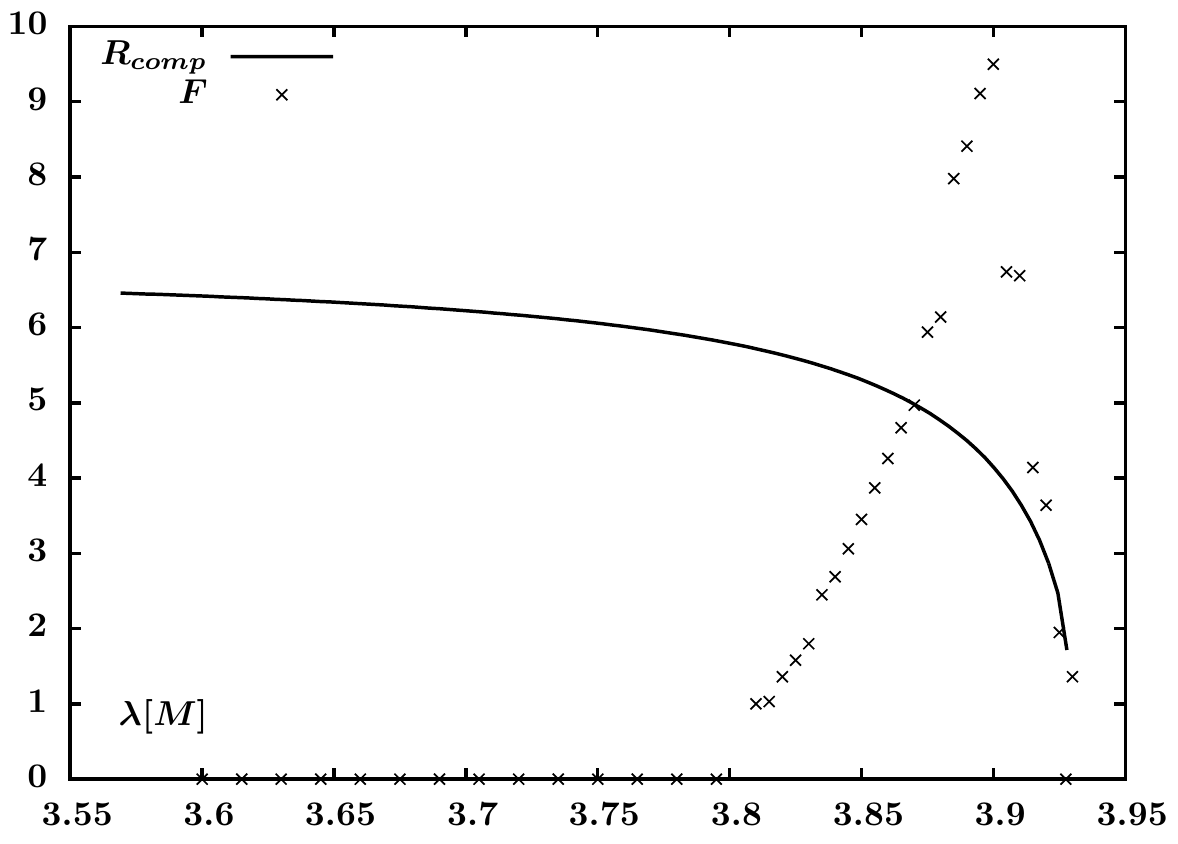}

\includegraphics[width=0.495\textwidth]{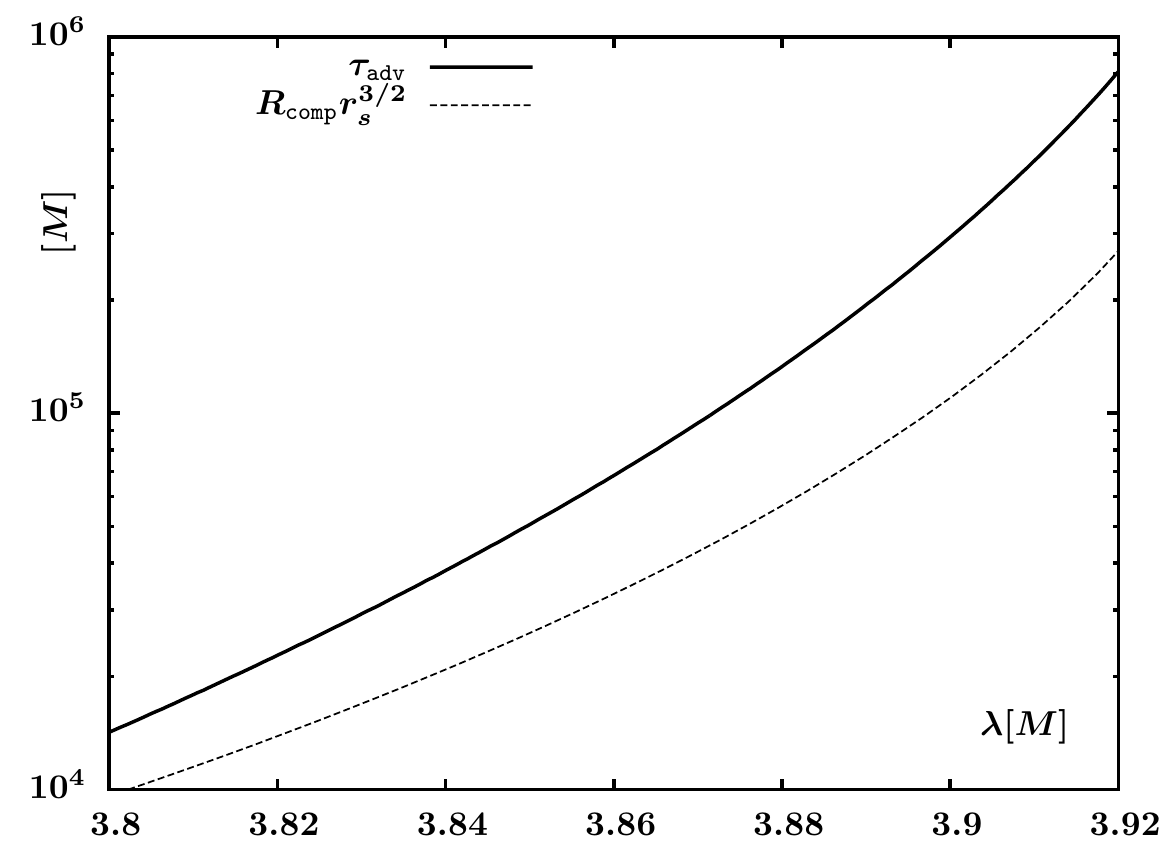}
\includegraphics[width=0.495\textwidth]{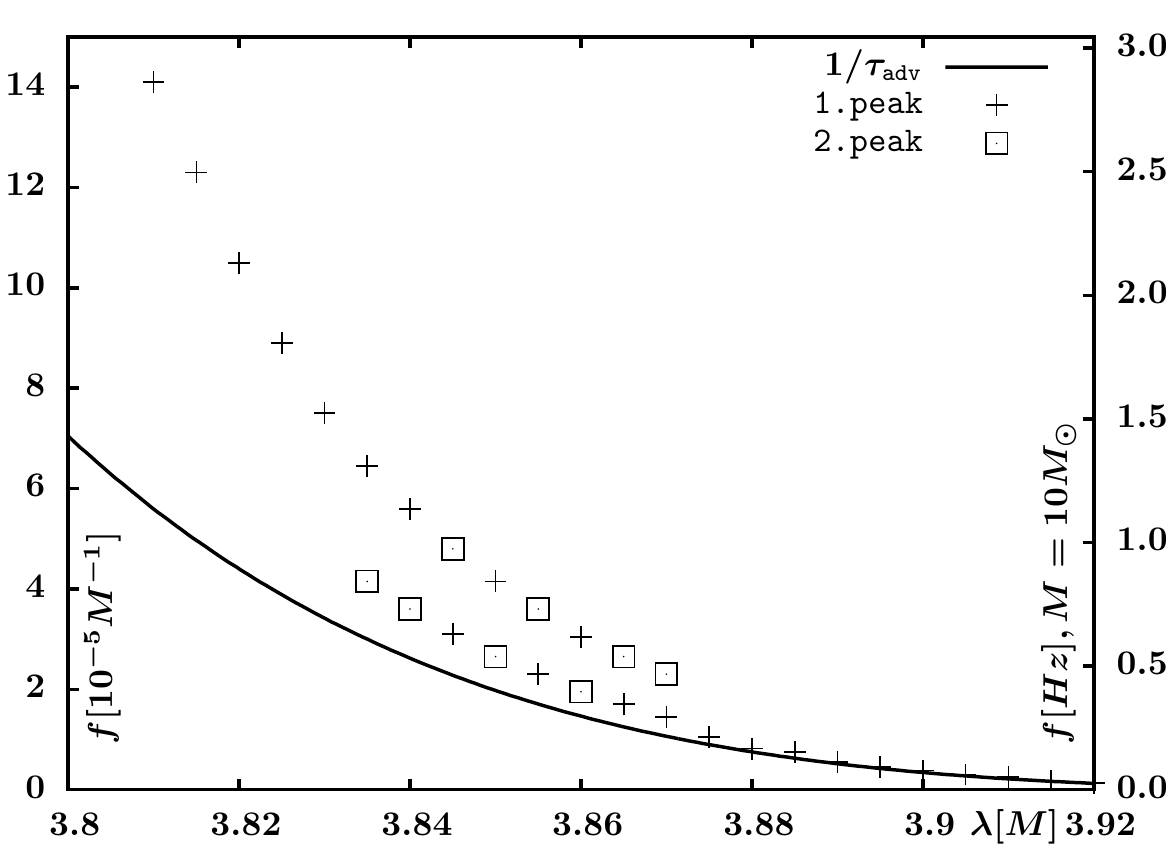} 
\caption
{First row: Dependence of the shock position $r_s$ (left), density compression ratio $R_{\rm comp}$ and accretion rate amplification factor $F=\dot{M}_{\rm max}/\dot{M}_{\rm min}$ (right) on the specific angular momentum $\lambda$ for $\epsilon=0.0001, \gamma=4/3$. The amplification factor $F$ is obtained for several $\lambda$ after the run of the simulations. Second row: Comparison of the advection time integrated from the shock position $\tau_{\rm adv}$ and its estimate from (\ref{nu_QPO}) (left) and the dependence of the highest peak (``plus" symbol) and the second highest peak (``square" symbol) of the spectrum together with the reciprocal value of the advection time on $\lambda$ (right).
\label{L_dep} \label{L_frek}} 
\end{figure*}

For our numerical study at first we adopt the parameters used by \cite{2002ApJ...577..880D} for Paczynski-Wiita potential, which are the energy $\epsilon=0.0001$ and the adiabatic index $\gamma=4/3$. The accretion rate is the Bondi accretion rate given by $\dot{M}=\lambda_c\frac{(GM)^2}{a_\infty^3}\rho_\infty$ (we skip $4\pi$ in definition (\ref{continuity})), where the coefficient $\lambda_c$ depends on the adiabatic index in the following way \citep{1952MNRAS.112..195B}
\begin{equation}
\lambda_c= \left( \frac{1}{2}\right)^{(\gamma+1)/(2(\gamma-1))} \left( \frac{5-3\gamma}{4} \right)^{-(5-3\gamma)/(2(\gamma-1))} .
\end{equation}
Because  the sound speed at infinity is given by
\begin{equation}
a_{\infty}=\sqrt{\epsilon(\gamma-1)}, \label{a_infty}
\end{equation} 
which comes from (\ref{E_cons}) for $r \to \infty$ and $u \to 0$, the density at infinity is related to the accretion rate by the relation
\begin{equation}
\rho_\infty = \frac{\dot{M}(\epsilon(\gamma-1))^{3/2}}{\lambda_c(GM)^2}.
\end{equation}

The value of the accretion rate  for all computations with different $\epsilon$ and $\gamma$ was chosen to be $\dot{M}=2\,000\,000$, which for parameters used in this section ($\gamma=4/3, \epsilon=0.0001$) gives $\rho_\infty=0.544$. Since the accreting gas is not source of gravitational field in our computations, the results scales just linearly with the value of density at infinity and the profiles of other variables are not influenced. 

For that particular values of parameters, the shock solution could be found for specific angular momentum in the range approx. $\lambda \sim(3.569\div 3.935)\,M$. We focus on the properties of the flow for different values of specific angular momentum in this range.

In Fig. \ref{fig_ha_hn} the shock solution is shown for two values of angular momentum $\lambda=3.6M$ and $\lambda=3.78M$. After a short initial time, when the shock is smeared a bit by the ZEUS algorithm, the solution settles down into the steady state and remain stationary until the computation is stopped for $t_{\rm max} = 4 \cdot 10^6\,M$. The plot depicts the radial course of the important dynamical quantities, the Mach number $\mathfrak{M}$, density $\rho$, inward velocity $u$, angular velocity $v_3$ and internal energy $e$. The points of intersection between the Mach number line and the line $y=1$ mark the sonic points and the shock position. {\color{black}For comparison, the shock-free outer branch, from which the shock solution diverges at $r_s$, is also depicted in the right panel of Fig. \ref{fig_ha_hn} (denoted by diamonds).}

The outer part of the profiles converges to the Bondi solution of spherical accretion with the sound speed at infinity given by (\ref{a_infty}). For parameters used in Fig. \ref{fig_ha_hn}, the Bondi sonic point is located at $r_{\rm Bondi}= 7506.66M$, which is only slightly farther out than the position of $r_{\rm out}$ in both cases ($r_{\rm in}=4.9M, r_{\rm out}=7485.04M$ for  $\lambda=3.6M$ and  $r_{\rm in}=4.39M, r_{\rm out}=7482.81M$  for  $\lambda=3.78M$). This corresponds to the fact, that the low value of specific angular momentum of the flow is not significant for the motion at long distances from the center.

Increasing the specific angular momentum further up, the behaviour of the system changes. The solution takes longer and longer time to settle down to the stationary solution and then around the value $\lambda=3.81$ the stationary state is not attained and the shock position starts to oscillate. In this case we can look at the time behaviour of the mass accretion rate through the inner boundary, because it is no longer a constant value. 

In Fig. \ref{fig_ic} the time variation of the ratio between the actual and initial value of the mass accretion rate, the normalized accretion rate $\dot{M}_n=\dot{M}(t)/\dot{M}(0)$, is shown for  $\lambda=3.83M$ in the top row, $\lambda=3.835M$ in the middle row and $\lambda=3.855M$ in the bottom row. The shock oscillates with a period and amplitude, which increases for increasing angular momentum. This is caused by the fact, that the shock position is placed further from the center, hence the time needed for the wave to arrive to the compact star is longer. Because the amplitude of the shock oscillation is bigger more mass falls down on the center during the phase, when the shock front is going down to the center, hence the amplitude of the accretion rate change is bigger. 

 The comparison of the time variation of the mass accretion rate and the shock position is depicted in Fig. \ref{fig_hv} for $\lambda=3.9M$. The normalized accretion rate is plotted with time delay corresponding to the advection time $\tau_{\rm adv}=293807M$, which is the time needed for the flow to fall down on the center from the initial shock position $r_s=888M$.  The shock position is searched in the grid from $r=8M$ up to higher values and is given as the zone coordinate, where the Mach number increases above one. The increase of the accretion rate corresponds to the increase of the shock position, which implies higher density at the inner boundary, and to the rapid downswing of the shock position, when the excess of matter is accreated quickly. For those cycles, when the shock position achieves significantly higher values then the initial shock position (around $1500M$), the peaks of accretion rate are more delayed corresponding to the higher advection time from that position.

\begin{figure*}
\includegraphics[width=0.495\textwidth]{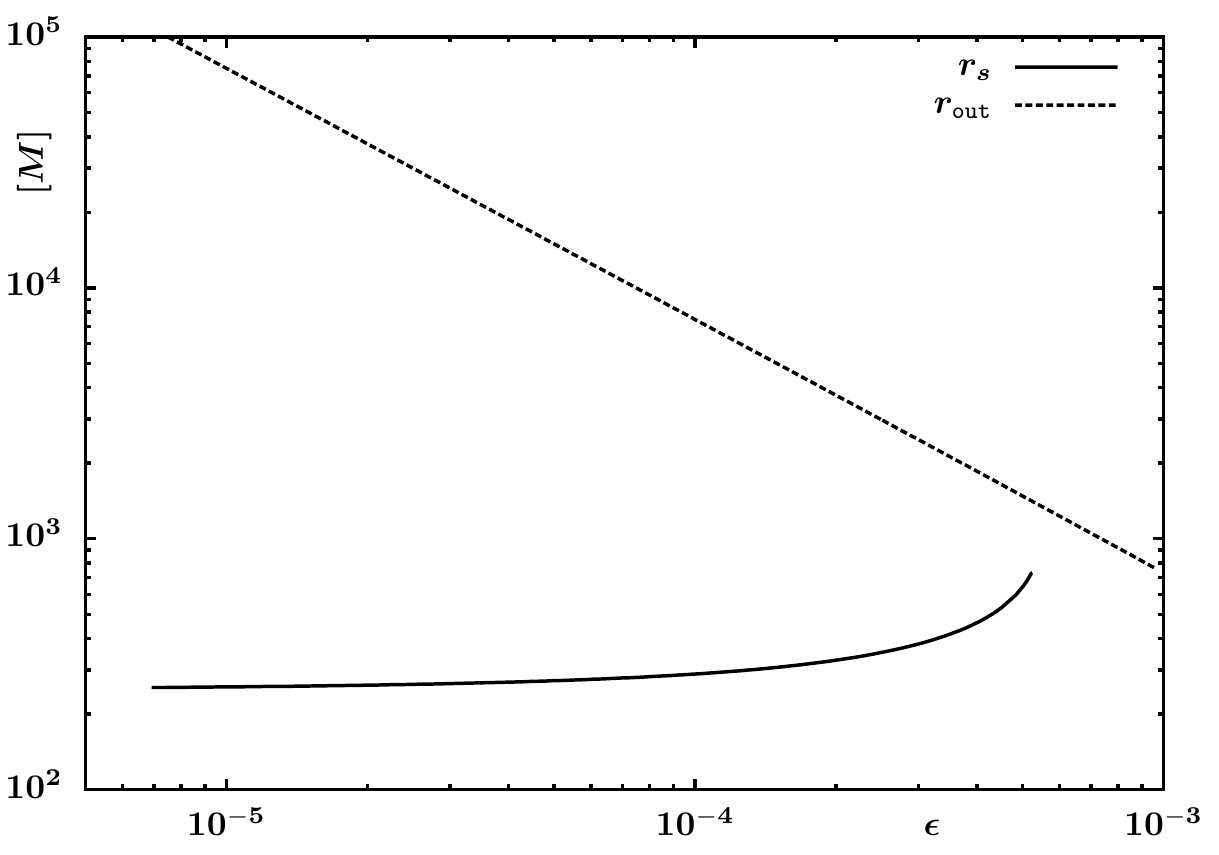}
\includegraphics[width=0.495\textwidth]{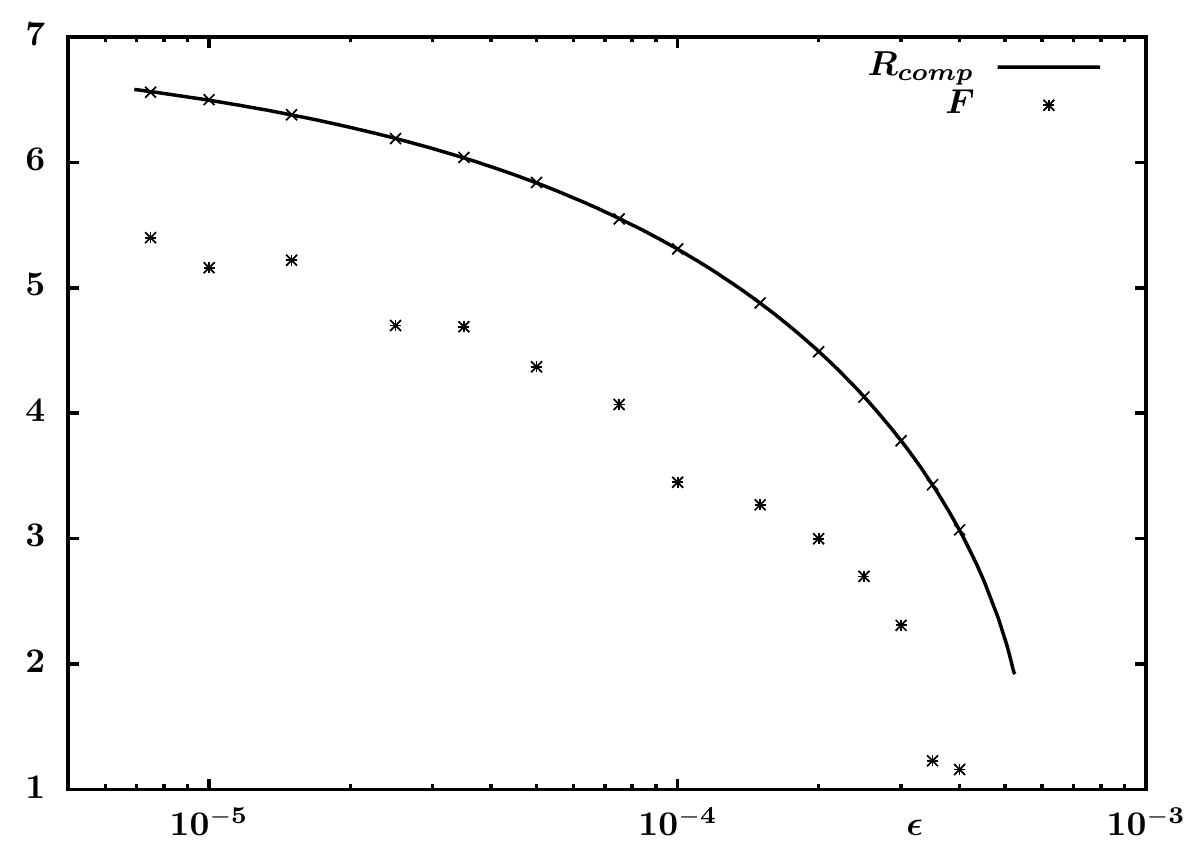}

\includegraphics[width=0.495\textwidth]{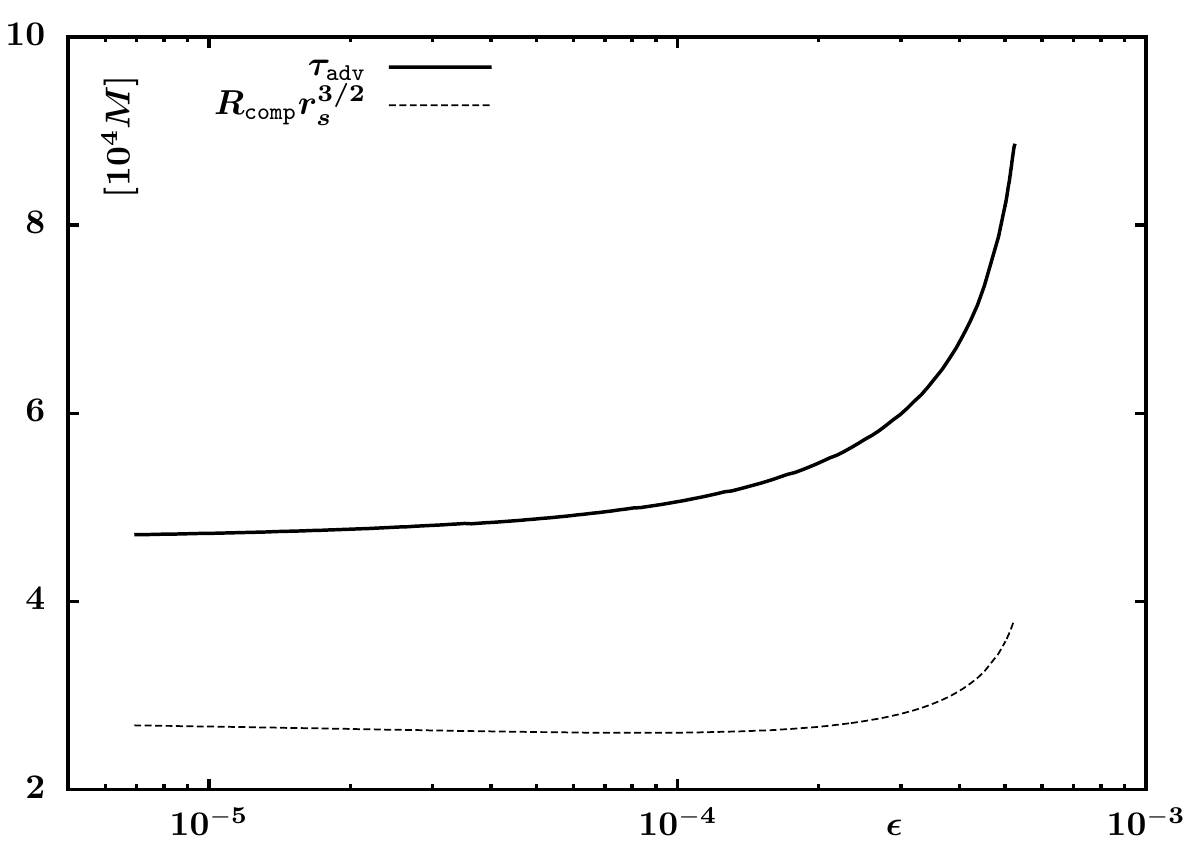}
\includegraphics[width=0.495\textwidth]{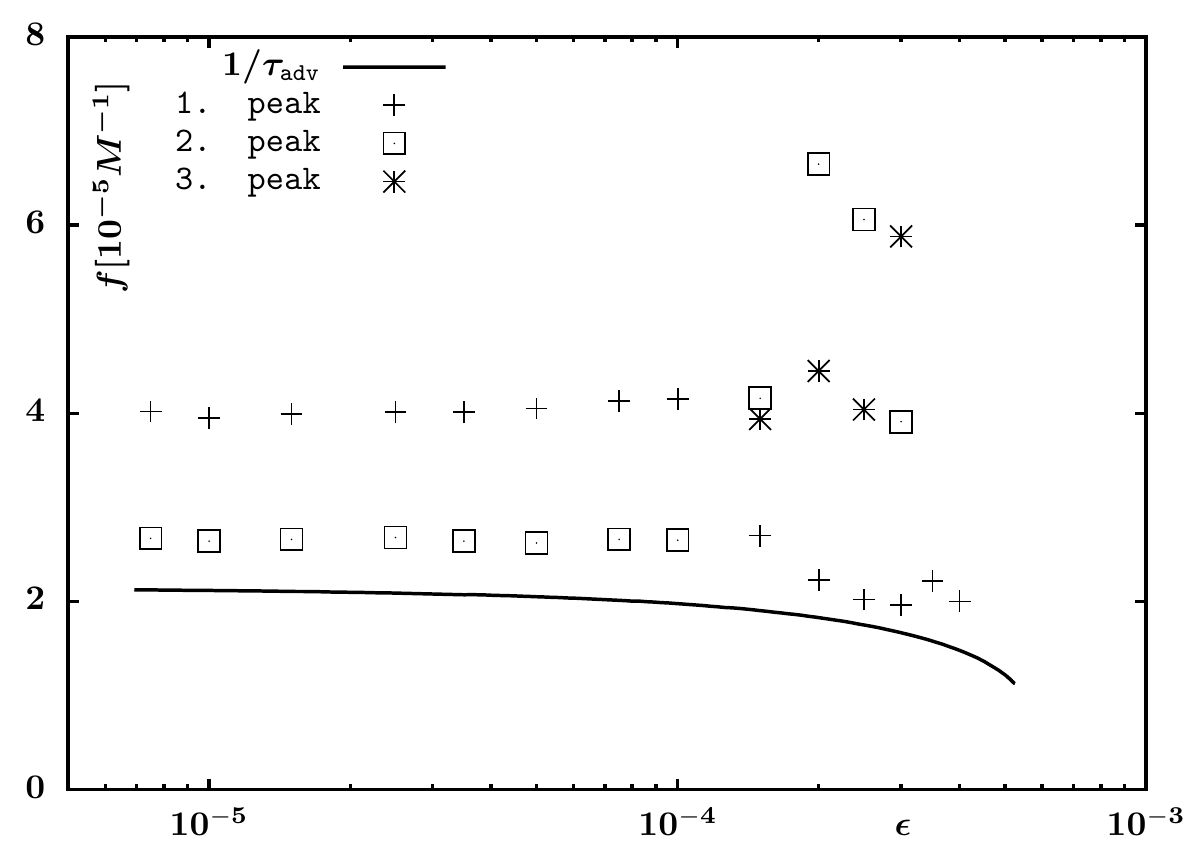} 
\caption
{First row: Dependence of the shock and outer critical point position (left), density compression ratio $R_{\rm comp}$ and accretion rate amplification factor $F=\dot{M}_{\rm max}/\dot{M}_{\rm min}$ (right) on the energy $\epsilon$ for $\lambda=3.85M$, $\gamma=4/3$. The amplification factor $F$ is obtained for several $\epsilon$ after the run of the simulations. Second row: Comparison of the advection time integrated from the shock position and its estimate from (\ref{nu_QPO}) (left) and the dependence of the highest (``plus" symbol), the second highest peak (``square" symbol) and third highest (``star" symbol) of the spectrum together with the reciprocal value of the advection time (right) on $\epsilon$.
\label{E_frek}}
\end{figure*}

From the time dependence of $\dot{M}$ we compute the spectrum by means of the fast fourier transform. The spectrum (plotted in the right column of Fig. \ref{fig_ic})  is obtained after the initial transition time. The frequency is expressed in the units $M^{-1}$, which can be scaled with the mass of the accreting object.\footnote{The time unit $1M$ equals to the time, which light needs to travel the half of the Schwarzschild radius of a black hole of mass M, hence for one solar mass $[t]=1M_\odot=\frac{GM_\odot}{c^3}=4.9255\cdot 10^{-6}s$.} For stellar black hole with typical mass $M=10M_\odot$ the upper x-axis shows the frequency expressed in $Hz$. For the used parameters $\gamma$ and $\epsilon$ and for $\lambda$ in the shock solution interval, the main frequency of the accretion rate oscillations is in the range $(3 \div 0.03) Hz$, even though a second ``bump" of the spectrum can be seen in the range $(30\div1)Hz$.

This behaviour of the shock would have observable imprints in the outgoing radiation flux from the system. If the radiation flux is linearly dependent on the accretion rate, then the change of the incoming flux by a factor 2 -- 10 with the mentioned frequency range would be detected, which may be a source of the measured  QPOs in some spectra.

The dependence of the shock position, the compression ratio $R_{\rm comp}$ and the amplification factor of the accretion rate $F=\dot{M}_{\rm max}/\dot{M}_{\rm min}$  on the specific angular momentum is plotted in Fig.~\ref{L_dep} in top row. The last quantity is set to zero for solution without the shock oscillations. 

The shock position $r_s$ and the compression ratio $R_{\rm comp}$ is obtained during the initial procedure, hence they are computed with higher accuracy than the grid resolution and more densely with the changing parameter. Their values anti-correlate with each other. Spectrum and amplification factor is obtained after the simulation evolves long enough, thus it is much more computer demanding to obtain their values and the accuracy depends on the resolution of the grid. 

The amplification factor of the accretion rate is obtained as the ratio of the maximal and minimal value of the accretion rate $F=\dot{M}_{\rm max}(t)/\dot{M}_{\rm min}(t)$ computed after the initial transient time. For lower values of $\lambda$ it correlates with the shock position and retains is maximum value $F_{\rm max}=9.5$ for $\lambda=3.9M$. For higher angular momentum it falls back down to one.

In Fig. \ref{L_frek} in bottom right panel we plot the dependence of the main frequency of mass accretion rate oscillations, which is obtained as the frequency of the main peak from the spectrum. For increasing angular momentum peaks at lower frequencies also occur, {\color{black} because the amplitude of different harmonics evolves with $\lambda$}, which can be seen in Fig. \ref{fig_ic} in the comparison of the spectra. For angular momentum $\lambda \in (3.835\div3.87)M$  two peaks of comparable amplitude can be found (visible for $\lambda=3.855M$ in the last spectrum in Fig. \ref{fig_ic}), which are both depicted in the plot, the highest peak is plotted with ``plus" symbol, while the ``square" symbol corresponds to the second highest peak. Although the switch between the highest and the second highest peak is not smooth with changing $\lambda$, together all the values form two very well defined branches, which both decrease with the increasing angular momentum.

\cite{2003ApJ...588L..89D}  suggested, that shock oscillations could appear due to the cooling of the flow. He proposed the dependence of the oscillation frequency on the postshock advection timescale $\tau_{\rm adv}$ in the form
\begin{equation}
\nu_{\rm QPO}=\frac{A}{R_{\rm comp}r_s^{3/2}}, \label{nu_QPO}
\end{equation}
with $A$ being the scaling constant, which should be obtained from the observations. However, the estimate of the advection timescale by the relation $\tau_{\rm adv} \approx R_{\rm comp}r_s^{3/2}$ is quite rough, because the total time depends not only on the postshock flow velocity but also on the profile of the inner branch of solution. In our case the profile of the inward velocity is known, hence we can integrate it to obtain the advection time directly. This is depicted in Fig. \ref{L_frek} on the left bottom plot together with the estimate $R_{\rm comp}r_s^{3/2}$.

Eventhough in our case cooling is not considered, {\color{black} our numerical simulations also show} oscillations of the shock position. In the right plot of Fig. \ref{L_frek} the obtained frequencies are compared with the reciprocal value of the advection time, which creates third branch of frequencies with smaller decline than the observed frequencies in the spectra. For higher specific angular momentum the reciprocal advection time agrees very well with the observed frequencies, while for smaller specific angular momentum, the main frequencies in the spectrum are higher with faster decline. On the right $y$-axis, the frequency in Hz is indicated for typical stellar black hole with mass $M=10M_\odot$.

In Fig. \ref{E_frek} the dependence of the important quantities on energy of the flow is shown for $\lambda=3.85M$ and $\gamma=4/3$. The shock solution is found for energy in the interval $\epsilon \in (7\cdot 10^{-6}, 5.3 \cdot 10^{-4})$.  In the first panel, the position of the shock and the outer critical  point is shown. In the log-log scale the outer critical point linearly decreases with increasing energy. Contrary, the shock position is increasing. Because the stable shock position is placed between $r_{\rm mid}$ and $r_{\rm out}$ \citep{2002ApJ...577..880D}, there is no shock solution for high energies. For small energies, the shock condition (\ref{shock}) is also not satisfied, hence the shock solution does not exist. Similar energy dependence of the parameter region with shock can be seen in Fig.~1 in \citep{2003MNRAS.343..443D} for the thin disc governed by hydrostatic equilibrium,  where the change of energy corresponds to vertical motion along the line $\lambda = {\rm const}$.  
In the right panel the profile of $R_{\rm comp}$ and $F$ is shown, which both decrease with increasing energy.  The amplification of the accretion rate during the oscillation is bigger for low energetic flows.

In the bottom left panel the integrated advection time $\tau_{\rm adv}$ is compared with the estimate by relation (\ref{nu_QPO}). The obtained advection time is longer than the estimate, which is caused by the fact, that the inward velocity right after the shock is quite high, but it is decreasing towards the center for higher shock position following the ``fish-shaped" inner branch of solution (as example see the inward velocity profile in the right panel of Fig. \ref{fig_ha_hn}).

Last panel of Fig. \ref{E_frek} shows the change of the main frequencies of the oscillation with the energy. For some energies, even three important {\color{black}harmonics} can be found. The comparison with the reciprocal value of the advection time shows, that the particular branch of frequencies follow similar tendency for the energy dependence, only being multiplied by scaling constant. For lower energies the higher branch of frequencies is dominant. For $\epsilon > 0.0003$ (last two symbols in the plot), the oscillations are dumped after the transient time and only stable shock solution with constant accretion rate is obtained. In the plot the frequency before the stabilization for these energies is shown. 

The frequency of the oscillation as well as the shock position does not depend strongly on the energy of the flow compared to the specific angular momentum dependence, eventhough the outer critical point position varies with energy 	through several orders of magnitude, while it is almost constant with changing angular momentum.

\begin{figure}
\includegraphics[width=0.48\textwidth]{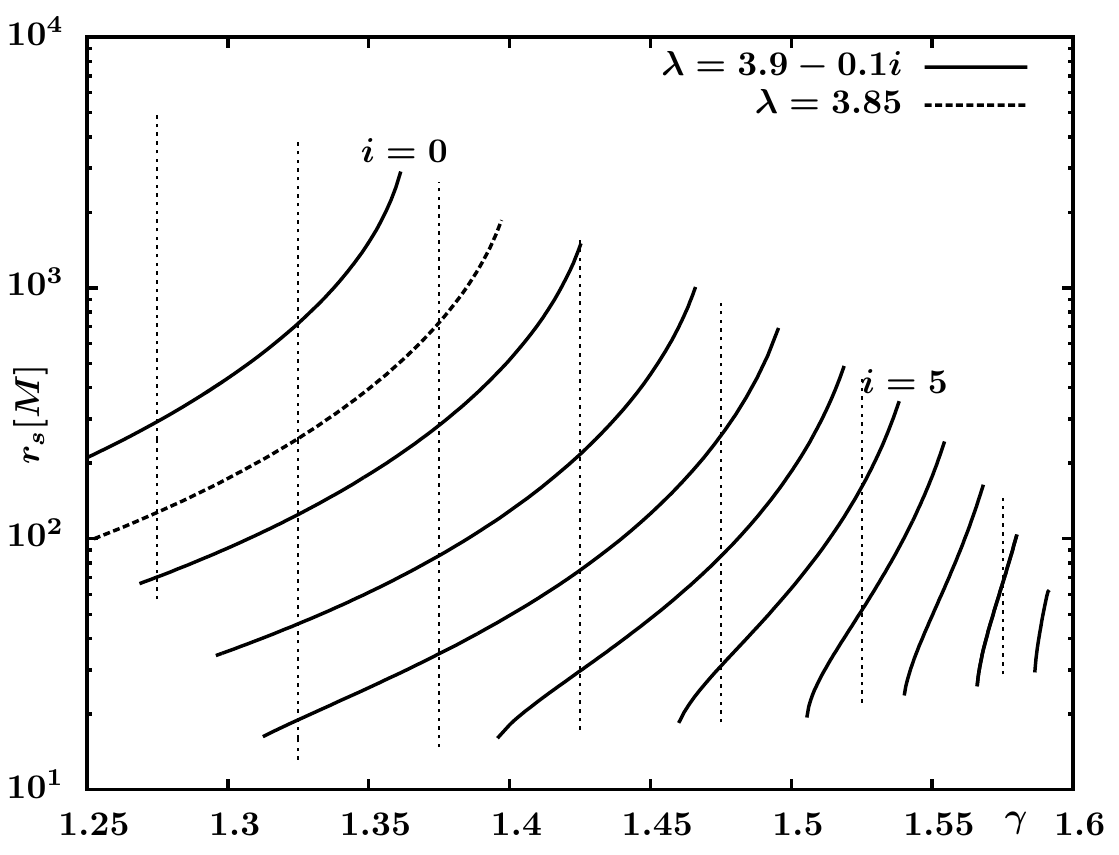}
\caption
{The dependence of the shock position on polytropic constant $\gamma$ is shown for 10 values of angular momentum from $\lambda=3.0M$ (solid line most right) with the step $0.1M$ to $\lambda=3.9M$ (solid line most left). The value $\lambda=3.85M$ used in Fig.~\ref{E_frek} is plotted by dashed line. The dotted lines represent the interval of possible shock position for given $\gamma$ and varying $\lambda$. \label{fig_gamma}
}
\end{figure}

The shock position and the shock existence interval for $\lambda$ strongly depends on the polytropic constant $\gamma$. This dependence is depicted in Fig.~\ref{fig_gamma} for $\epsilon=0.0001$ and several values of angular momentum $\lambda_i=3.9M-0.1Mi, i=0..9$ by solid lines and $\lambda=3.85M$ by dashed line. The vertical dotted lines represents the possible shock positions for given $\gamma$ and varying $\lambda$.  The minimal and maximal shock positions ${r_s}_{\rm min}$ and ${r_s}_{\rm max}$ get closer for $\gamma$ increasing from the ultrarelativistic flow $\gamma=4/3$ to the nonrelativistc limit $\gamma=5/3$, where only negligible shock interval for small angular momentum exists.

\subsection{Simulations of the shocked flow with changing $\lambda$} \label{loop}

Another type of variability of the outgoing radiation from a  quasi-spherical system with low angular momentum apart from the shock oscillation for certain values of $\lambda$ discussed above could be caused by a slow change of the angular momentum due to variation of the conditions around the accreting object. For stellar binary systems this variation could be caused by the eccentricity of the orbital trajectories and thus would happen on orbital time scale. For supermassive black holes in the AGNs the modulation could happen due to stellar disruption or different cloud accretion with more or less stochastic appearance or due to stellar wind coming from the orbiting stars with eccentric orbits.

\begin{figure*}
\includegraphics[width=0.495\textwidth]{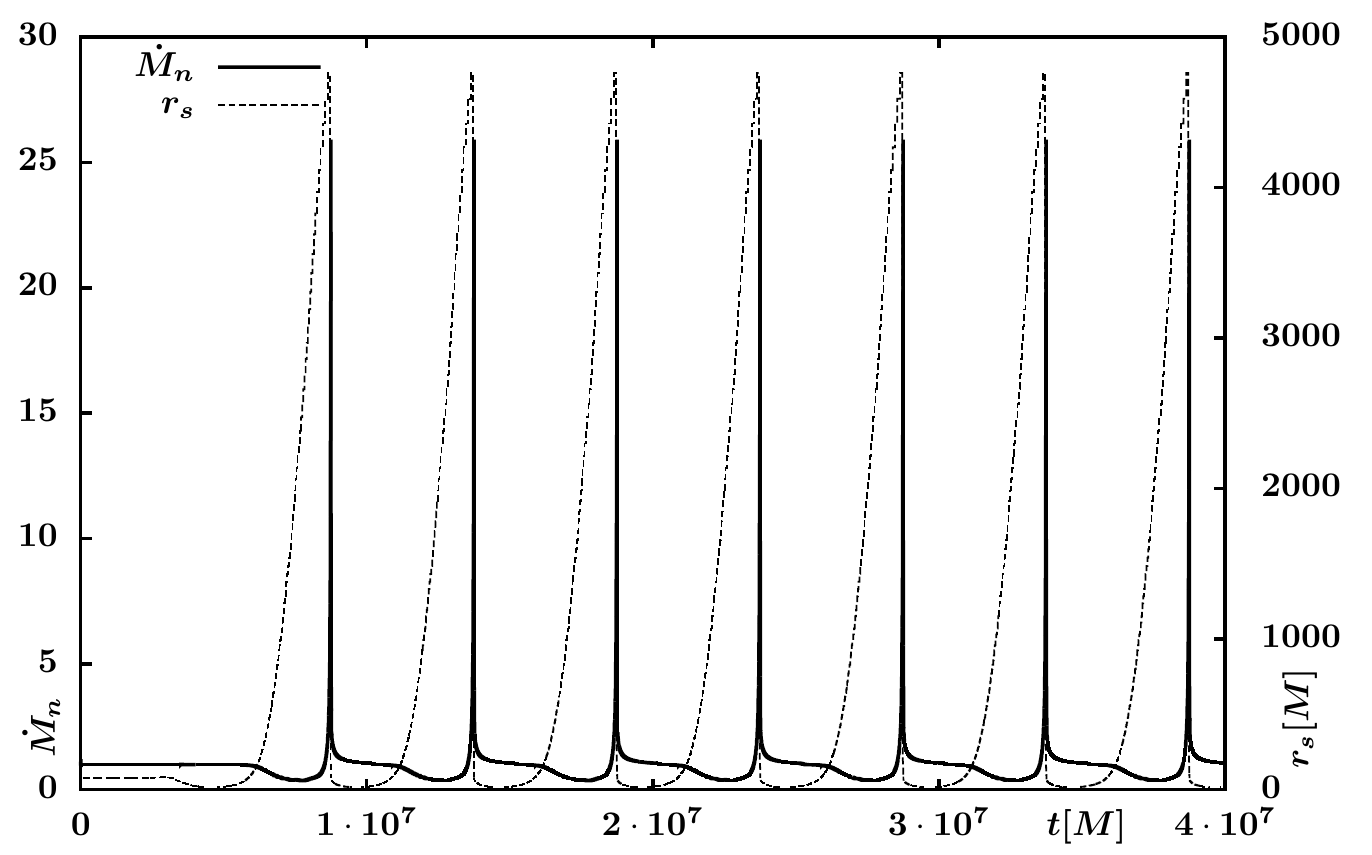}
\includegraphics[width=0.495\textwidth]{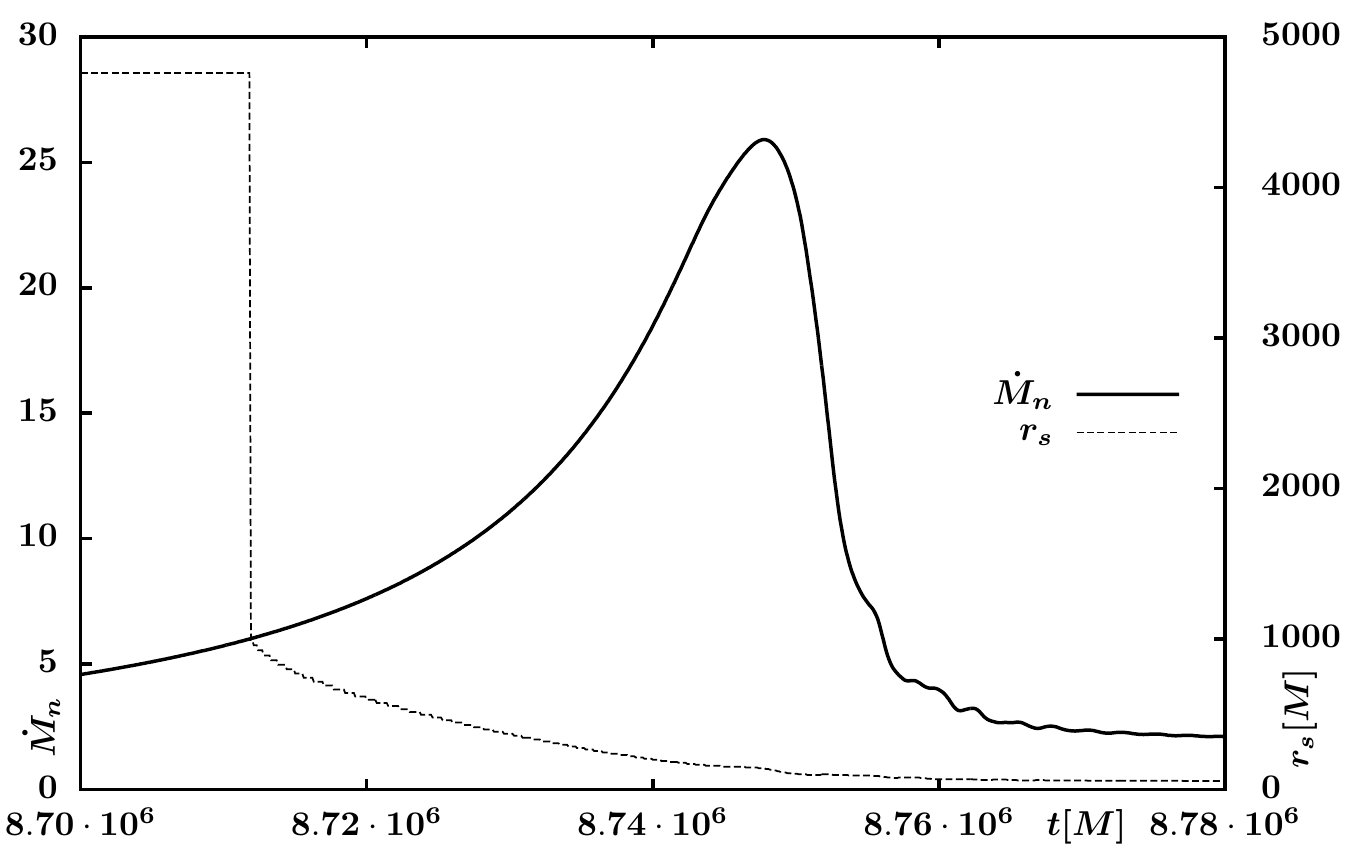}
\caption
{Time variation of normalized accretion rate $\dot{M}_n=\dot{M}/\dot{M_0}$ through the inner boundary (solid line) and the position of the shock $r_{\rm s}$ (dashed line) for $\epsilon=0.0001$, $\gamma=4/3$, $\lambda_0=3.76M$, $\Delta \lambda = -0.18M$ and $T_v=5\cdot 10^6M$. On the right the zoom of the plot around the first peak is shown. \label{fig_qd}
}
\end{figure*}

In this section we study numerically the behaviour of the shocked flow, when the specific angular momentum of the material incoming through the outer boundary changes periodically. This is done in the following way. For $\gamma=4/3$ and $\epsilon=0.0001$ we prescribe the initial condition in the same way like in the preceding section with some angular momentum $\lambda_0$. During the computation the boundary values of density, internal energy and radial velocity in the two ghost zones at the outer boundary is being kept constant given by the initial evaluation. The value of the angular velocity is changed by the prescription
\begin{equation}
v_3(t)=\lambda(t)/r, \qquad \lambda(t) = \lambda_0 + \Delta \lambda \sin\left(\frac{2\pi t}{T_v}\right),
\end{equation}
where $\Delta\lambda$ is the amplitude of the angular momentum variation and $T_v$ is the period of the variation.

For this task the parameters of the grid, $\Delta \lambda$ and $T_v$ should be carefully set. During the evolution of the flow with angular momentum going to lower values the position of the inner sonic point get closers to the marginal value $r_{\rm in}=4M$. If this point is very close to the inner boundary of the grid (let say, if the sonic point comes to second or third zone in the grid), then the computation with the outgoing inner boundary condition is not stable, because the flow is not supersonic enough and some spurious waves reflect on the inner boundary. Therefore we choose the inner boundary to be $r_{\rm ib}=3.9M$ and the grid  dense enough in this region. On the other hand at the outer boundary the matter with changing angular momentum should flow into the grid from the ghost zones. For doing so, it has to have enough time considering the size of the last zone of the grid (in our logarithmic scale it is  roughly $300\div1000M$) and the inward velocity of the flow at the outer boundary. In Fig.~\ref{fig_ha_hn} the inward velocity for $r=10^5M$ is seen to be close to $10^{-4}$, which means, that travelling through the last zone of grid would last several millions of $M$. With respect to this time the period of the variation $T_v$ should be much longer, otherwise the angular momentum perturbation would not pass through the outer part of the grid inside. Because of the fine grid near the inner boundary the minimal time step is given by the CFL stability condition and is approx. $10^{-1}M$, computation of such long time scale is thus quite demanding. Therefore we used  two logarithmic scale for the radial grid for these simulations, which is more dense at the inner boundary and not so sparse at the outer boundary. We also choose smaller $r_{\rm ob}=20\,000M$, which is still {\color{black} more than} two times further from the center than the position {\color{black} of the outer sonic point}. The inward velocity is one order of magnitude higher at this radius. With this settings, the acceptable period $T_v$ is in the order of millions of $M$, which is possible to compute in reasonable CPU time.

In physical time units $10^7M$ corresponds to $\sim 500s$ for $M=10M_\odot$ (stellar black hole) and to $\sim 570$ days for $M=10^6M_\odot$ (for Sgr~A* $M \sim 4.3 \cdot 10^6M_\odot  \rightarrow 10^7M \sim 2 450 d$). This is still by several orders shorter than the typical observed orbital scale for the low-mass X-ray binaries (e.g. \cite{2014arXiv1406.7262G} reported orbital period about 30 days for two black hole candidates GRS 1915+105 and IGR J17091-3624), but it is quite in agreement with the orbital periods of the closest stars near the supermassive black holes in AGNs (e.g. S0-102, the closest star to Sgr~A*, has orbital period $T_o=11,5yr=4200d=1.7 \cdot 10^7M$ \citep{Meyer05102012}).

\begin{figure*}
\includegraphics[height=0.33\textwidth]{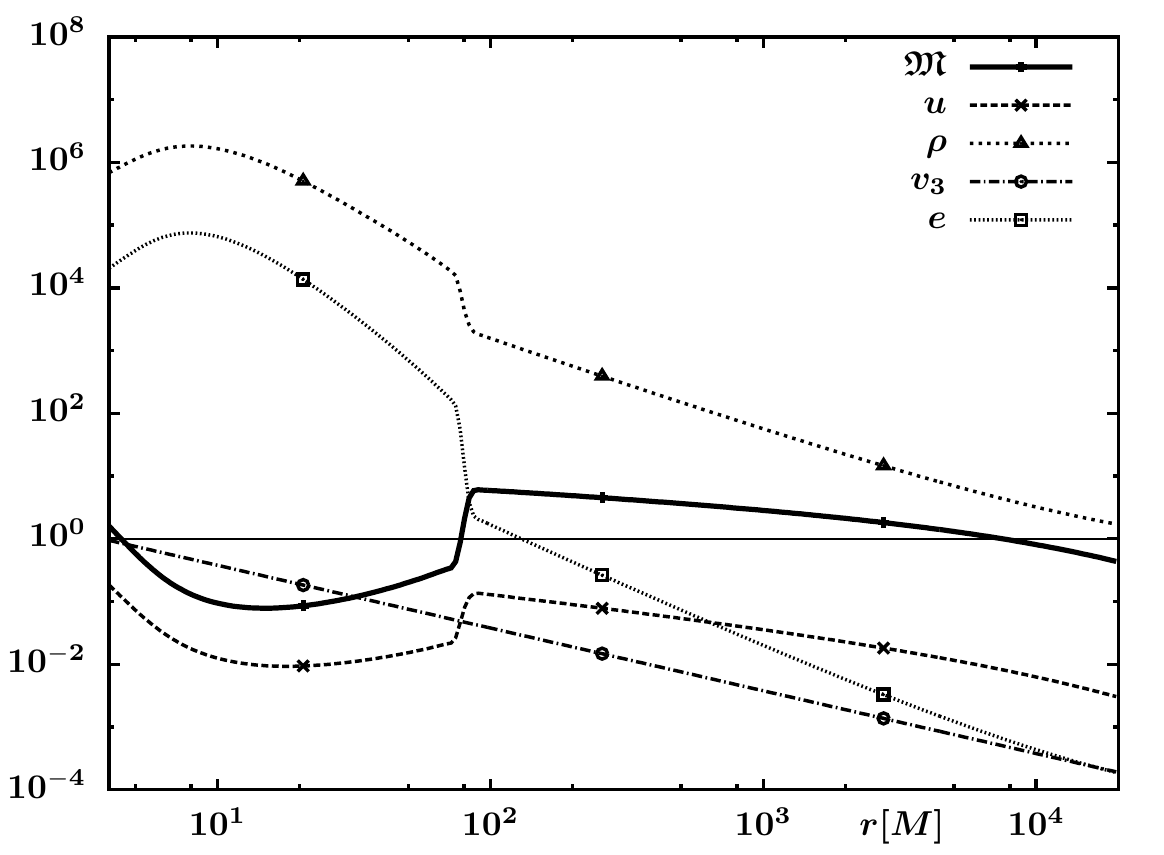}
\includegraphics[height=0.33\textwidth]{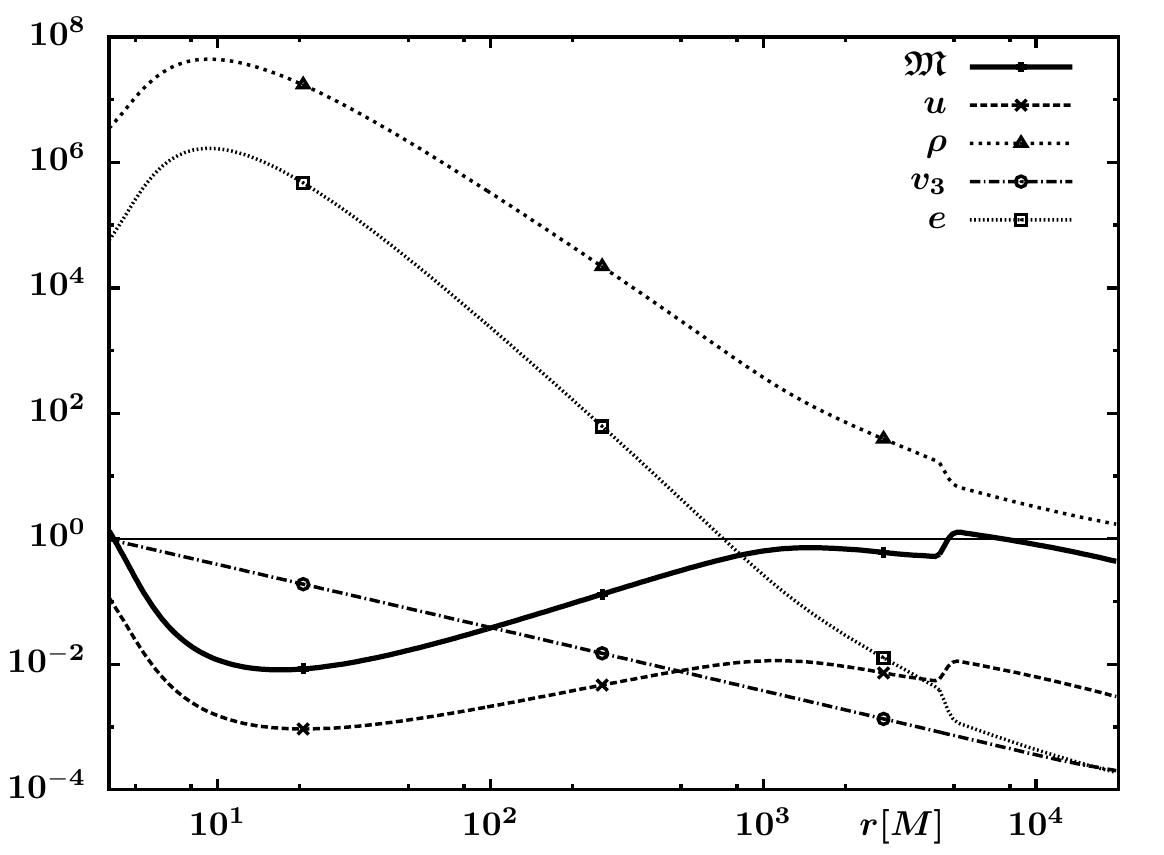}

\includegraphics[height=0.33\textwidth]{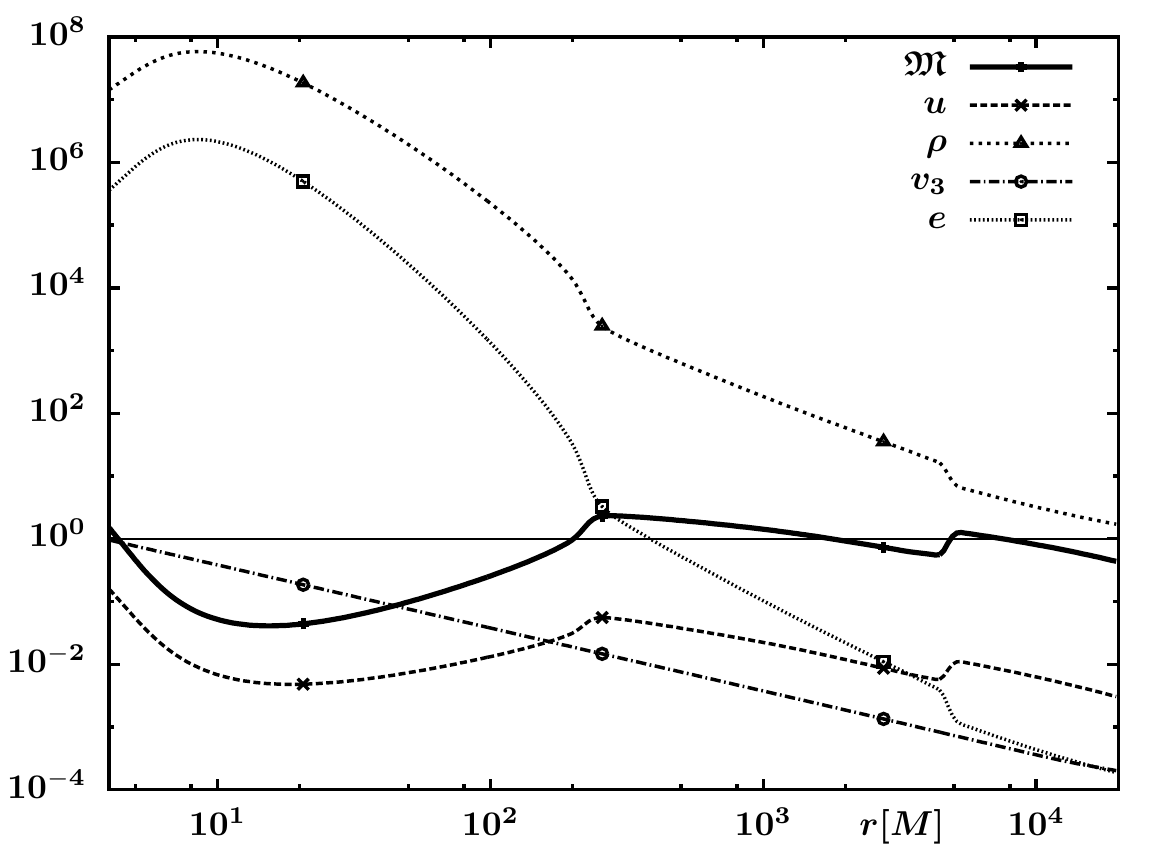}
\includegraphics[height=0.33\textwidth]{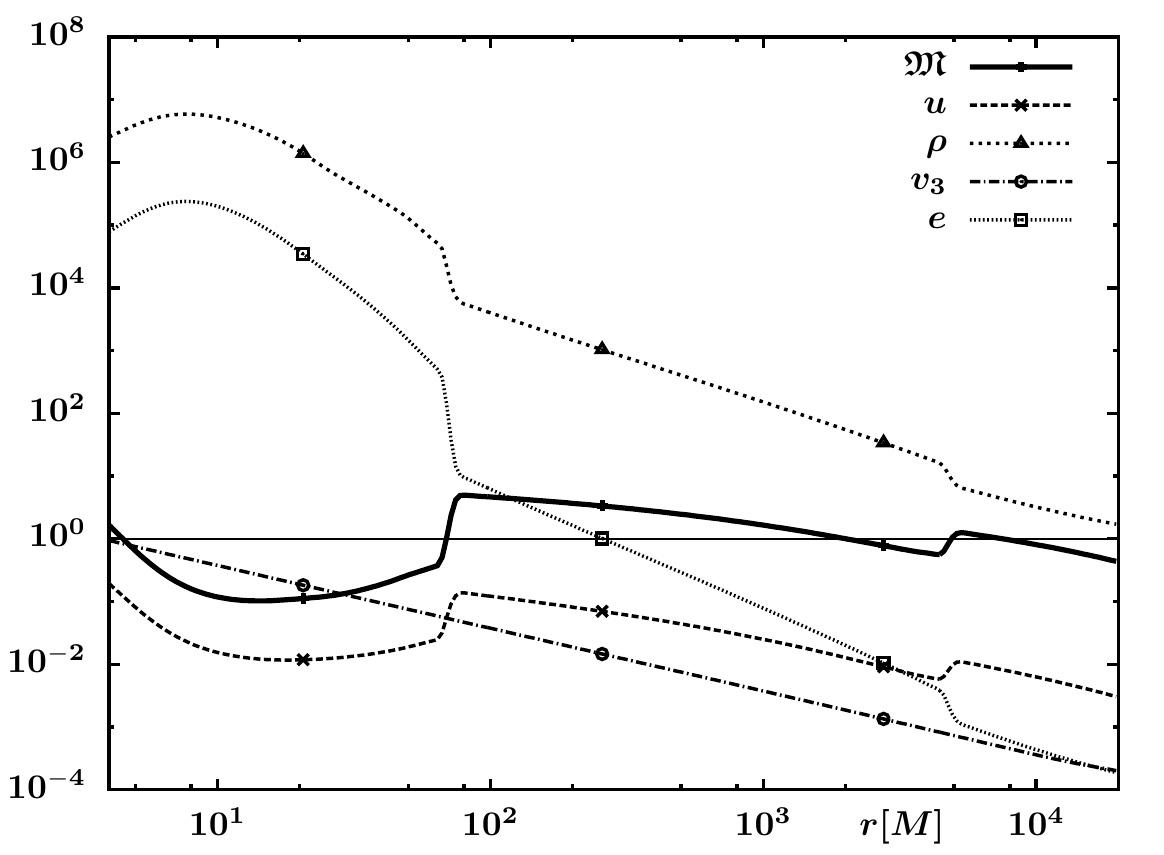}
\caption
{Snapshot of the accretion flow profiles at times $t_1=2\cdot 10^4M$ (top left), $t_2=8.68 \cdot 10^6M$ (top right), $t_3=8.74 \cdot 10^6M$ (bottom left) and $t_4=8.76 \cdot 10^6M$ (bottom right). The meaning of the plot is the same as in Fig. \ref{fig_ha_hn}. The parameters are $\gamma=4/3$, $\epsilon=0.0001$, $\lambda_0=3.76M$, $\Delta\lambda=-0.18M$, $T_v=5\cdot 10^6M$.\label{fig_qd_hdf}
}
\end{figure*}

In Fig. \ref{fig_qd} the time variation of the accretion rate (solid line) and the shock position (dashed line) for $\lambda_0=3.76M$, $\Delta \lambda = -0.18M$ and $T_v=5\cdot 10^6M$ is shown. On the left the whole integration interval of $4 \cdot 10^7M$ is depicted, which covers 8 full periods of the specific angular momentum variation and 7 sharp peaks could be seen in the accretion rate. This is caused by the fact, that the variation amplitude $\Delta \lambda$ was chosen to have negative value, which means, that the angular momentum decreases first and then increases. The shock position reacts to this variation with some time delay caused by the travelling time of the angular momentum perturbation from the outer boundary to the inner part of the flow. With this delay the shock position decreases and increases to very high radius. During this interval the density in the inner part increases significantly, correspondingly to the stationary solution with higher angular momentum and this higher profile of density expands to much higher radii (several snapshots of the flow profile are given in Fig. \ref{fig_qd_hdf}). Thus there is a lot of matter under the shock front. This enhancement of the density travels with another time delay ($\sim \tau_{\rm adv}$) to the inner boundary and increase the accretion rate. When the angular momentum decreases again, the shock front moves back down to the center and this extra amount of matter is accreted during a short period of time (note that the time step between the two bottom panels in Fig. \ref{fig_qd_hdf} is only $\Delta t=2\cdot 10^4M$). The amplification of the accretion rate is in this case much higher than for the oscillating shock with constant angular momentum.

A zoom of the accretion rate and shock position during the first peak is depicted on the right plot of Fig. \ref{fig_qd}.  The half-width of the peak $t_h = 1.7\cdot 10^4M$ is much shorter time than the variation amplitude of the angular momentum, $t_h/T_v = 0.0034$.  Hence, even for long lasting smooth variation of the angular momentum at the outer boundary, the accretion rate shows very narrow and high peaks. Thus, {\color{black} similar mechanism proceeding in the part of the accretion flow with low angular momentum} could be responsible for the bright flares observed in some sources. 

Comparing the accretion rate zoom and shock position zoom, we can also determine the delay between the fast shock drop down and the peak of the accretion rate at the inner boundary, which is in this case $t_d=3.5 \cdot 10^4M$. The shape of the peak fits quite well the observed bright flare from Sgr~A* \citep{2012ApJ...759...95N}, which is asymmetric with slow rise and sharp decline. However, in Sgr~A* the observed flares have the timescale in the order of thousands of seconds, which corresponds to some hundreds of $M$ and is still much shorter than these results (the one mentioned in \citet{2012ApJ...759...95N} lasted $5.6\,ks \sim 260\,M$).

\begin{figure*}
\includegraphics[width=0.495\textwidth]{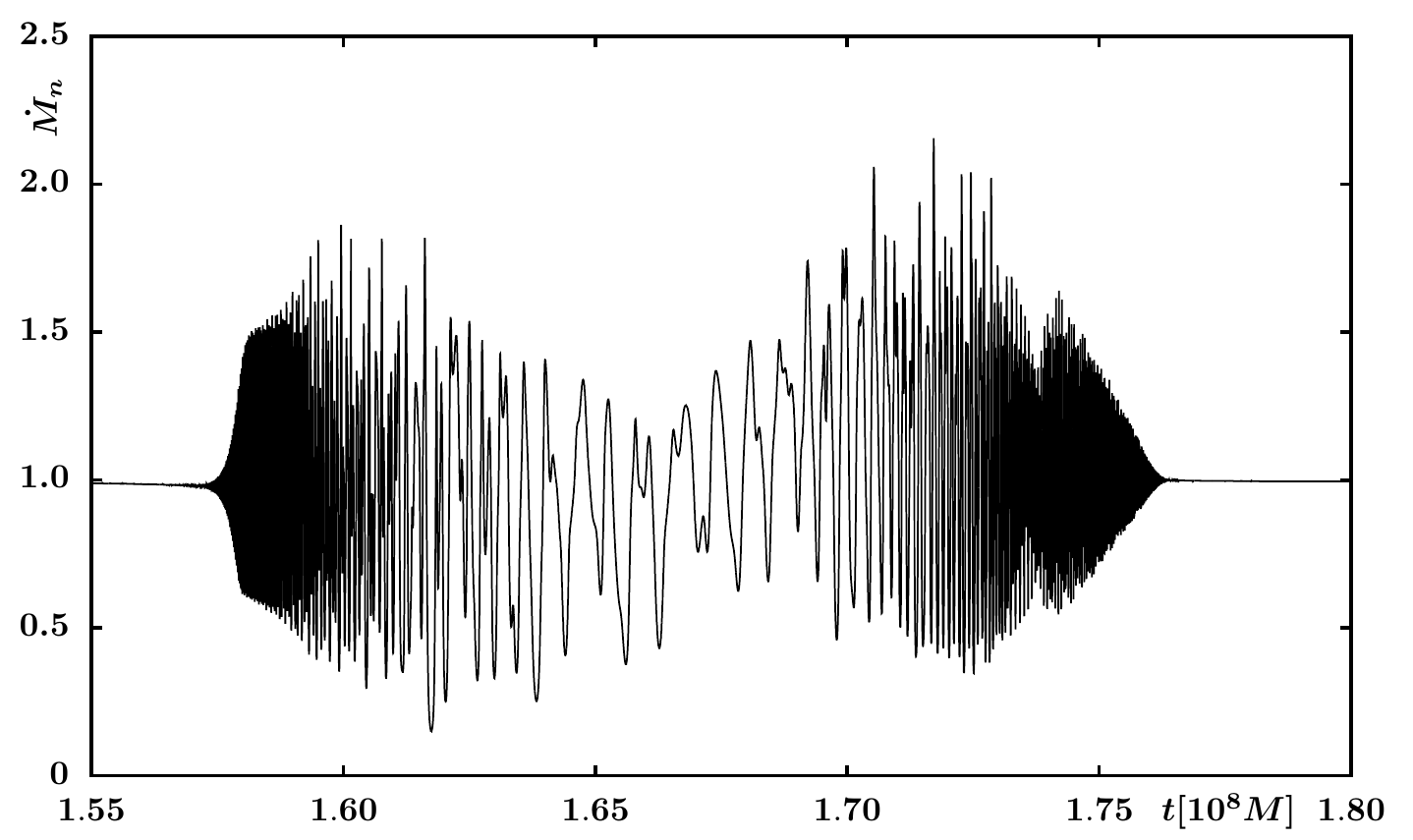}
\includegraphics[width=0.495\textwidth]{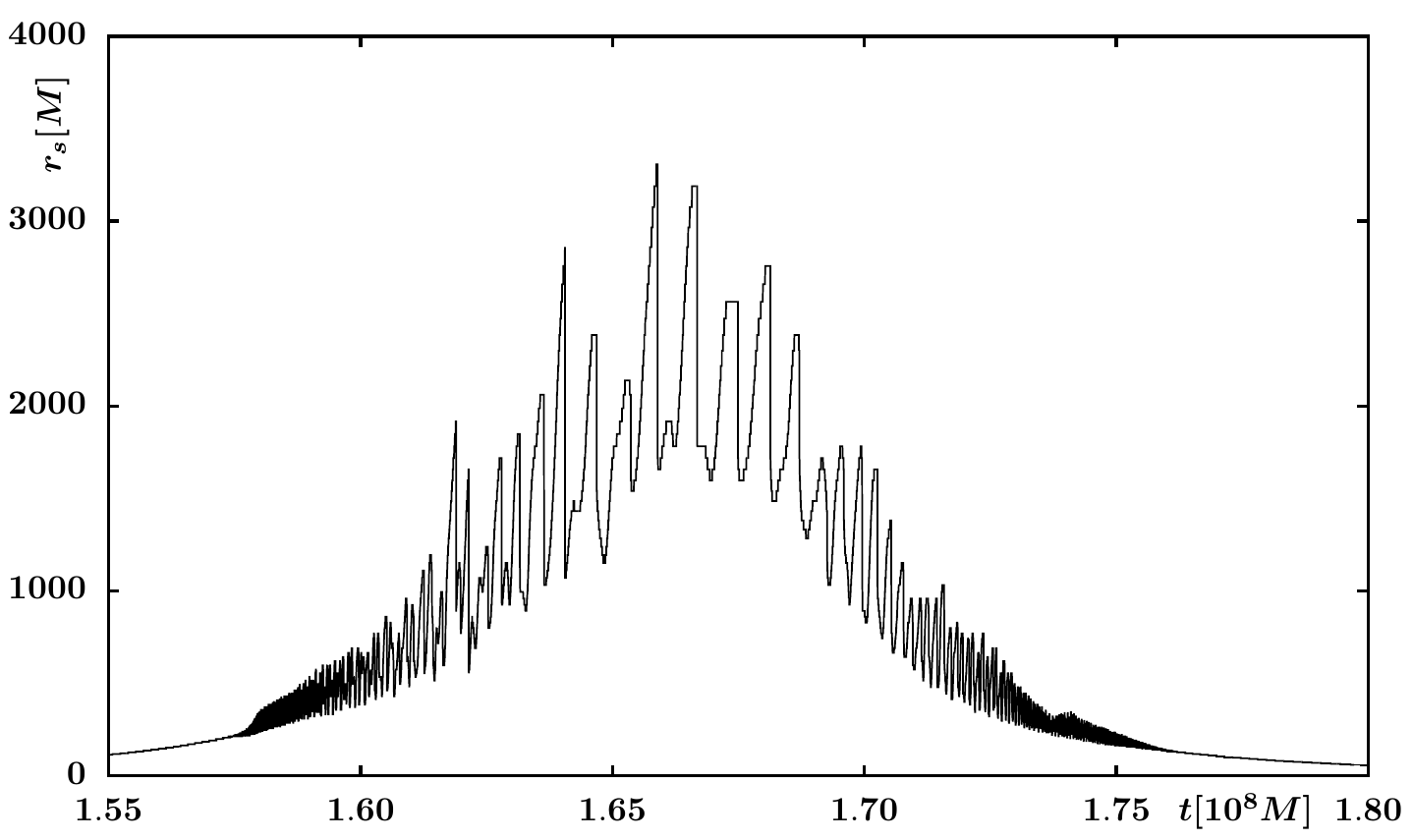}
\caption
{Time variation of normalized accretion rate $\dot{M}_n=\dot{M}/\dot{M_0}$ through the inner boundary (left) and the position of the shock $r_{\rm s}$ (right) for $\epsilon=0.0001$, $\gamma=4/3$, $\lambda_0=3.76M$, $\Delta \lambda = 0.16M$ and $T_v=5\cdot 10^7M$.  \label{fig_qu}
}
\end{figure*}

The scaling of the accretion rate peak half-width and amplitude with the variational period $T_v$ is given in Table~\ref{tab-1}. For longer periods the amplitude of the peak decreases and after the peak also oscillatory behaviour appears, which is given by the shock front oscillation for certain range of $\lambda$. For very long variational period and small amplitude there is not one single peak in the accretion rate, but it rather oscillates with gradually changing frequency and amplitude corresponding to the actual value of angular momentum. This behaviour is achieved for ten times longer period $T_v=5\cdot10^7M$ and somewhat smaller amplitude $\Delta\lambda=0.16M$ than in the preceding example (Fig.~\ref{fig_qu}).

\begin{table}
\begin{tabular}{c|c|c|c|c}
$T_v[10^6M]$ & $t_h [10^3M]$   & $t_h/T_v$   & $\dot{M_n}_{\rm max}$  & $t_w[M]$\\
$5$    & $6$   &  0.0012   &  51.4 & 74.5  \\
$10$    & $17$    &  0.0017   & 22.4 & 72.7\\
$20$    & $94$    &   0.0047  & 10.4 &  69.7 \\
$50$    & $230$    &  0.0046   &  5.9 &  69.2 \\
\end{tabular}
\caption{Half-width of the accretion rate peak $t_h$ and the maximal amplitude of the peak $\dot{M_n}_{\rm max}$ for different period of angular momentum $T_v$ for $\epsilon=0.0001$, $\gamma=4/3$, $\lambda_0=3.5M$, $\Delta\lambda=0.45$.  Width of the small accretion rate peak  $t_w$, which is the time interval, when the normalized accretion rate is higher than 1.2.
\label{tab-1}
}
\end{table}

\cite{2012NewA...17..254D} suggested a possible hysteresis loop for the sequence of steady solutions with smoothly varying leading parameter (specific angular momentum in this case). For the set of parameters for which multiple critical points exist, the solution either can pass through the outer sonic point only or it can form the shock and go through both sonic points. The density in the inner region varies significantly between these two options. The choice between them can depend on the history of the evolution. When the parameter enters the multicritical region from below, where only the solution through the outer critical point exists, it maintains similar properties, hence no shock forms for all the values of $\lambda$ in the multicritical region. After $\lambda$ comes to the region, where only solution through inner sonic point exists, abrupt change of the solution properties happens. Decreasing $\lambda$ back to the multicritical region (where the inner branch could not extend to infinity), the shock forms keeping the inner branch of solution in the inner accretion region tying it up with the outer branch far away from the center. 

In our case we do not study the sequence of steady solutions, but instead we change the boundary conditions with slow change of $\lambda$ as described above. As has been said before, the multicritical region with shock solution for $\epsilon=0.0001$ and $\gamma=4/3$ is $\lambda\in(3.569;3.935)M$. The hysteresis effect is thus present for  $\lambda_0$ and $\Delta\lambda$ such that values smaller than $3.57$ and higher than $3.93$ are achieved. There has to be sufficient time for the angular momentum change to travel through the grid. {\color{black} Also,}  when the solution switches from the outer branch to the inner branch, a wave of sudden change from supersonic to subsonic motion is sent up to higher radii, where the flow should be subsonic everywhere. When this wave arrives at the outer boundary, it bounced back, because the boundary is set to inflow boundary condition with the prescribed boundary values obtained from the outer branch of solution. Such bouncing wave can destroy the numerical solution. However, if the change of the parameter is fast enough, then the decreasing value of parameter in the outer region causes the shock formation at high radius and this wave is smeared before it hits the outer boundary. In that case the solution maintains its numerical accuracy.

With these restraints in mind we compute the problem with $\lambda_0=3.74M$, $\Delta\lambda=0.18M$, $T_v=10^7M$ and $r_{\rm ob}=20\,000\,M$.
Our results show that the proposed hysteresis effect applies also in the dynamical evolution of the accretion flow with slowly changing specific angular momentum at the outer boundary. In Fig.~\ref{fig_qp} the snapshots of the evolution are shown. In time  $t_1=2.01\cdot 10^7M$ the shock is placed at the minimal possible radius ${r_s}_{\rm min}$ for this $\gamma$ and $\epsilon$. Then the angular momentum in the inner region decreases below 3.58 and the flow passes through the outer sonic point only (second snapshot at $t_2=2.015 \cdot 10^7M$). For increasing $\lambda$ this type of solution remains, only the Mach number  profile is approaching one close to $r=4M$. The outer solution holds until specific angular momentum in the inner region  reaches 3.93, when the Mach number touches the line $y=1$ (third snapshot at $2.57 \cdot 10^7M$). At that moment the inner solution forms in the inner region, which generates a wave changing the outer solution into the inner one going upwards (fourth snapshot at $t_4=2.575 \cdot 10^7M$, where the front of the wave is placed at $r=340M$). However, in the outer part of accretion flow, the angular momentum decreases back into the shock existence interval, so the outer branch is preserved and shock appears at very high radius (fifth snapshot at $t_5=2.695 \cdot 10^7M$). For decreasing $\lambda$, the shock moves downwards again and the cycles repeats (sixth snapshot at $t_6=2.795 \cdot 10^7M$). {\color{black} Hence, during our simulation the shock in the flow creates by itself from the shock-free state and disappears repeatedly due to the change of the angular momentum.}

The accretion rate variation for this hysteresis loop is plotted in  Fig.~\ref{fig_qp} in the first row. The sharp drop of the accretion rate corresponds to the reshaping of the flow between the outer and inner branch, when the density at the inner boundary decreases significantly (could be seen at  $t_4=2.575 \cdot 10^7M$). The increase is then given by the modification of the density and velocity profile due to the shock appearance and its motion downwards (because in steady state the accretion rate is constant and the same for all parameters and the higher density for the inner branch is compensated by smaller inward velocity). The oscillation at the end of the peak are caused by the oscillating shock position for that $\lambda$, for which we observed oscillation in the preceding section.

The second smaller and sharper peak is created when the shock is accreted for $\lambda<3.58$ and only the outer branch remains (between the first and second snapshot). At the top right panel the zoom of this peak is shown with much smaller time step for the output (solid line). With the fine time resolution we can see, that the peak is almost as high as the first one and the width of the peak $\Delta t \approx 60M$. This value is quite similar to the mentioned shorter time scale of flares of Sgr~A* and is related to the smallest possible stable shock position, which is ${r_s}_{\rm min} = 12M$ for the used parameters. 

As we have shown earlier the position of the shock does not depend strongly on the energy of the flow (Fig.~\ref{E_frek}), but it does depend on the value of $\gamma$ (Fig.~\ref{fig_gamma}). For the comparison, we computed similar problem with $\epsilon = 3.3\cdot 10^{-6}$, which according to \cite{2012NewA...17..254D} should be appropriate value for accretion of wind from nearby stars by Sgr~A*, and for two values of polytropic constant $\gamma^2=1.5$ and $\gamma^3=1.6$. The shock existence interval is $\lambda^2 \in (3.36;3.80)$ and  $\lambda^3 \in (3.03;3.26)$. Similar peaks caused by the same phenomenon  are present in the accretion rate profile and the zoom of them is plotted in the top right panel of  Fig.~\ref{fig_qp} by dashed line for $\dot{M}^2$ and dotted line for $\dot{M}^3$ with the same time resolution. The time axis $\tilde{t}=t-t_0$ is shifted by $t_0^1=10168175M$, $t_0^2=17778500M$ and $t_0^3=8951542M$ in order to directly compare the peaks. Also the hight of the second and third peak is scaled to the right $y$-axis for the same reason. Because for higher $\gamma$ the shock existence interval is smaller and the last stable position of the shock is placed further away from the center (${r_s}^2_{\rm min} = 19M$ and ${r_s}^3_{\rm min} = 33M$), the two additional peaks are broader but their maximal amplitude is smaller.  However, if we take into account, that not only the accretion rate but also the significant difference of the density between the outer and shock solution contribute to the outgoing radiation, the amplitude of the corresponding peak could be even higher. This feature will be investigated in the future. 

\begin{figure}
\includegraphics[width=0.48\textwidth]{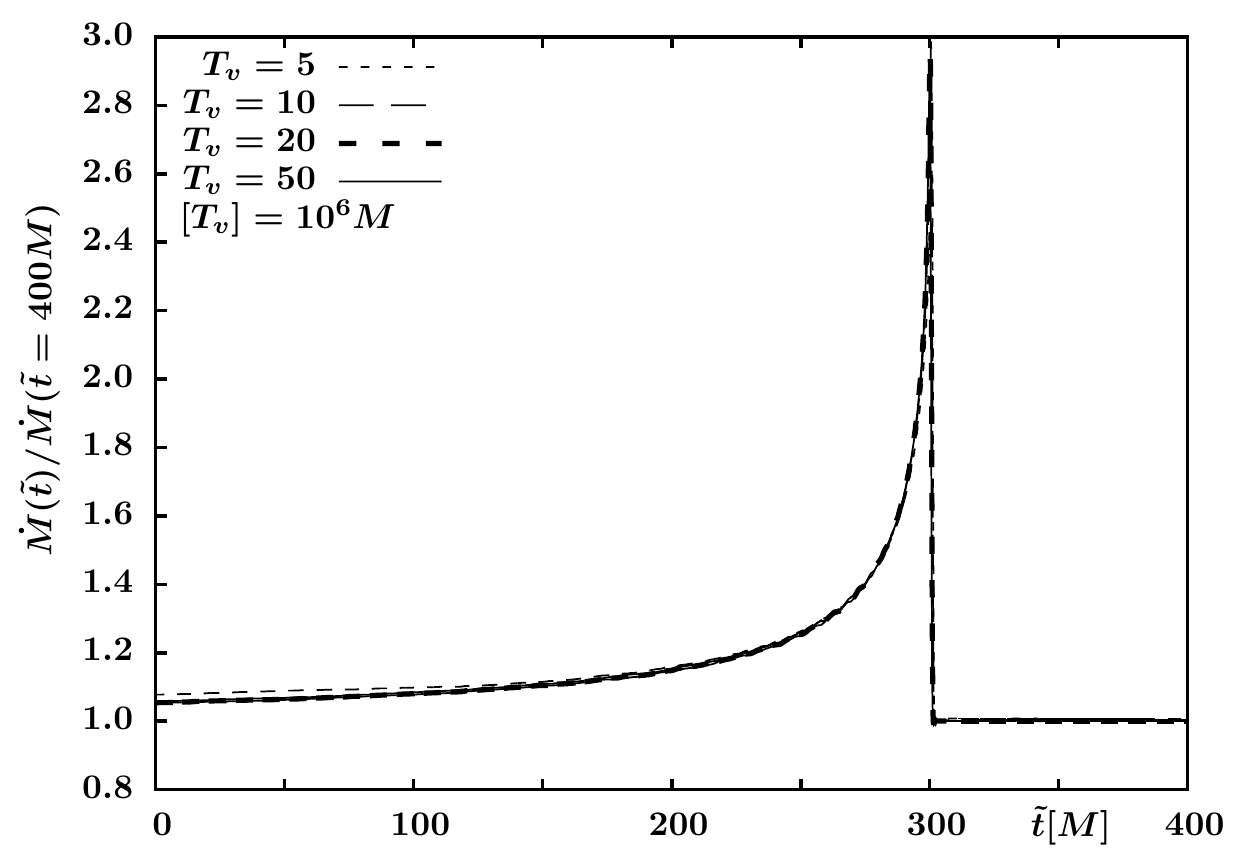}
\caption
{ Small peaks in accretion rate caused by the switch between the shock and outer solution for the simulations listed in Table~\ref{tab-1}.  The time axis is shifted so that the maximum of the peak is placed at $\tilde{t}=300M$. The accretion rate values are normalized with respect to the value of accretion rate after the decline of the peak at $\tilde{t}=400M$. The shape of the peak is independent on the variational period $T_v$. \label{fig_rk}
}
\end{figure}


The time scale of these peaks is in good agreement with the measured Sgr~A* flares, ranging between $50M\sim1000s$ to $250M\sim5300s$. Because of the very slow rise at the begging of the peak, the moment when such increase could be observable depends also on other processes in the accretion region. {\color{black} However, this mechanism can induce such flare only if some part of the accretion flow posseses low angular momentum.}

In Table~\ref{tab-1} also the time scale $t_w$ of the small peaks for different $T_v$ is given for constant $\gamma$ and $\epsilon$. In accordance with the fact, that ${r_s}_{\rm min}$ is the same, the width of the peak is almost constant and the profiles of the peaks are basically identical for all values of the variational period (plotted in Fig.~\ref{fig_rk}). Hence we can conclude that in case that the flow switches from the shock solution into the outer branch of solution, peak in accretion rate occurs, whose time scale is given  by ${r_s}_{\rm min}$  and accordingly mainly by the polytropic constant $\gamma$.

\begin{figure*}
\includegraphics[width=0.495\textwidth]{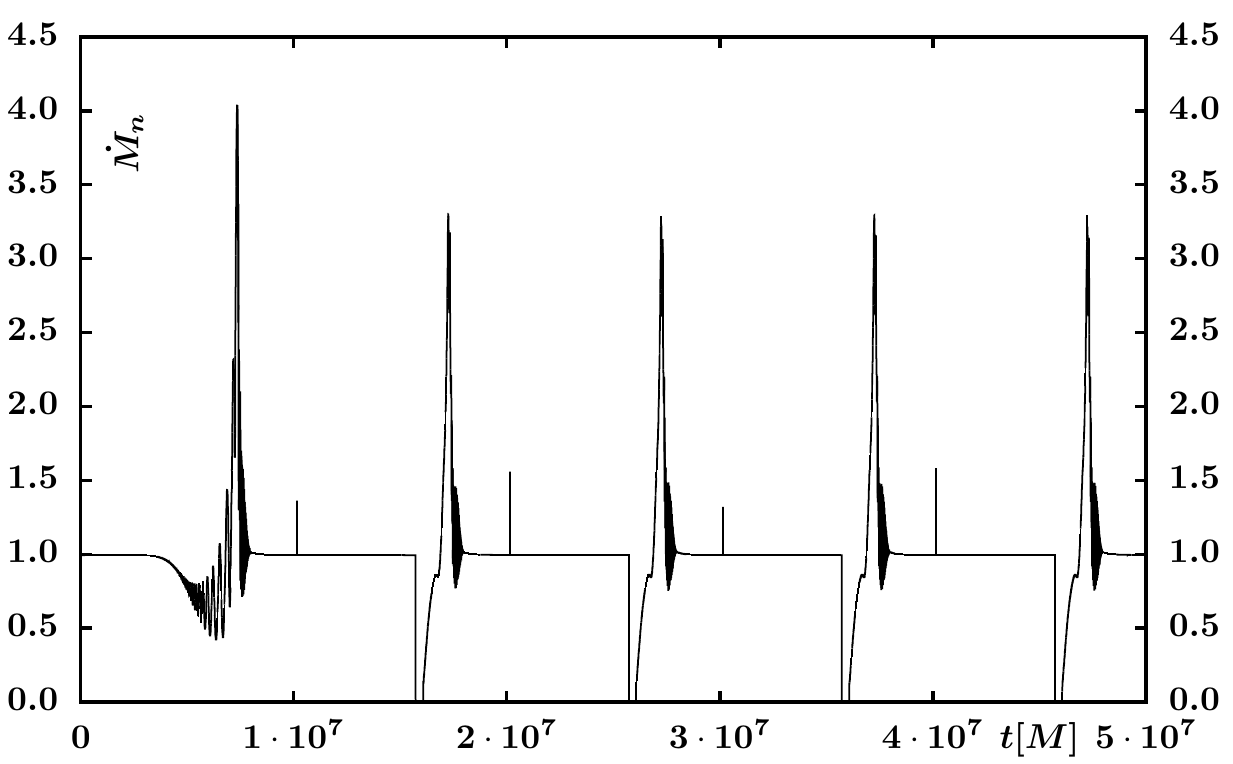}
\includegraphics[width=0.495\textwidth]{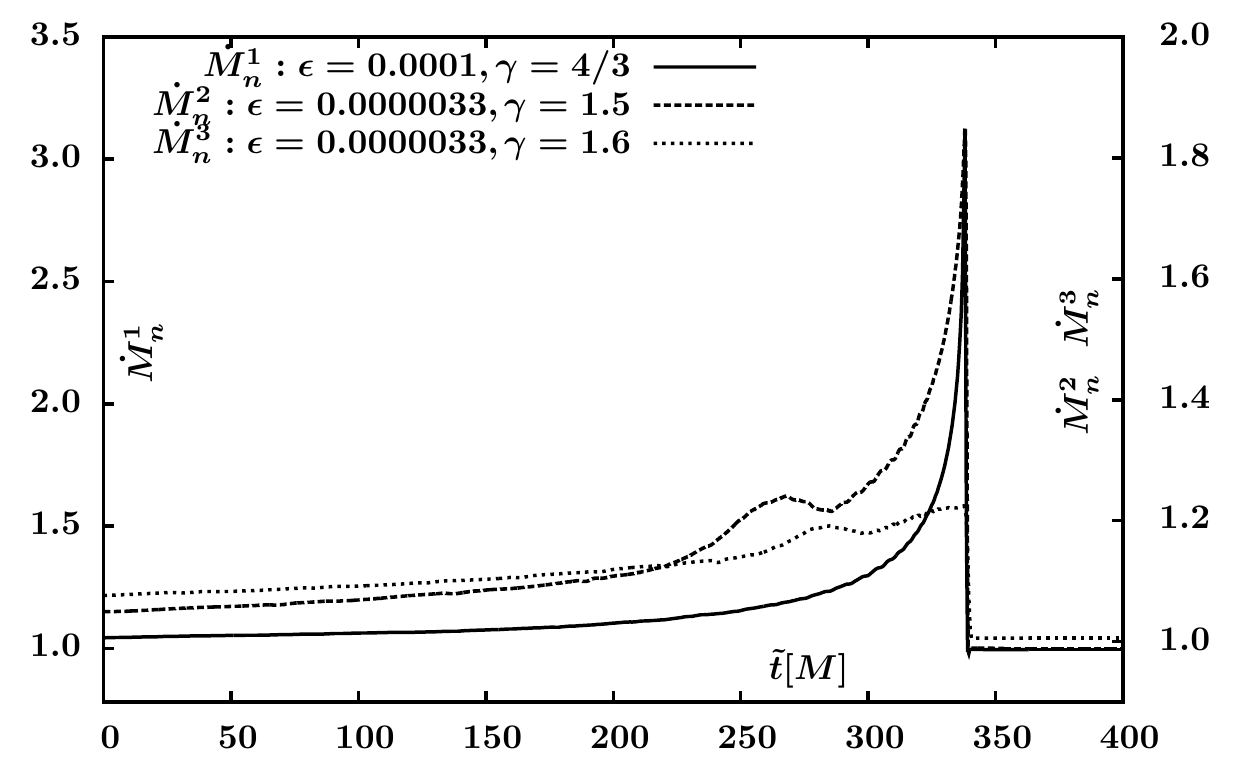}

\includegraphics[width=0.495\textwidth]{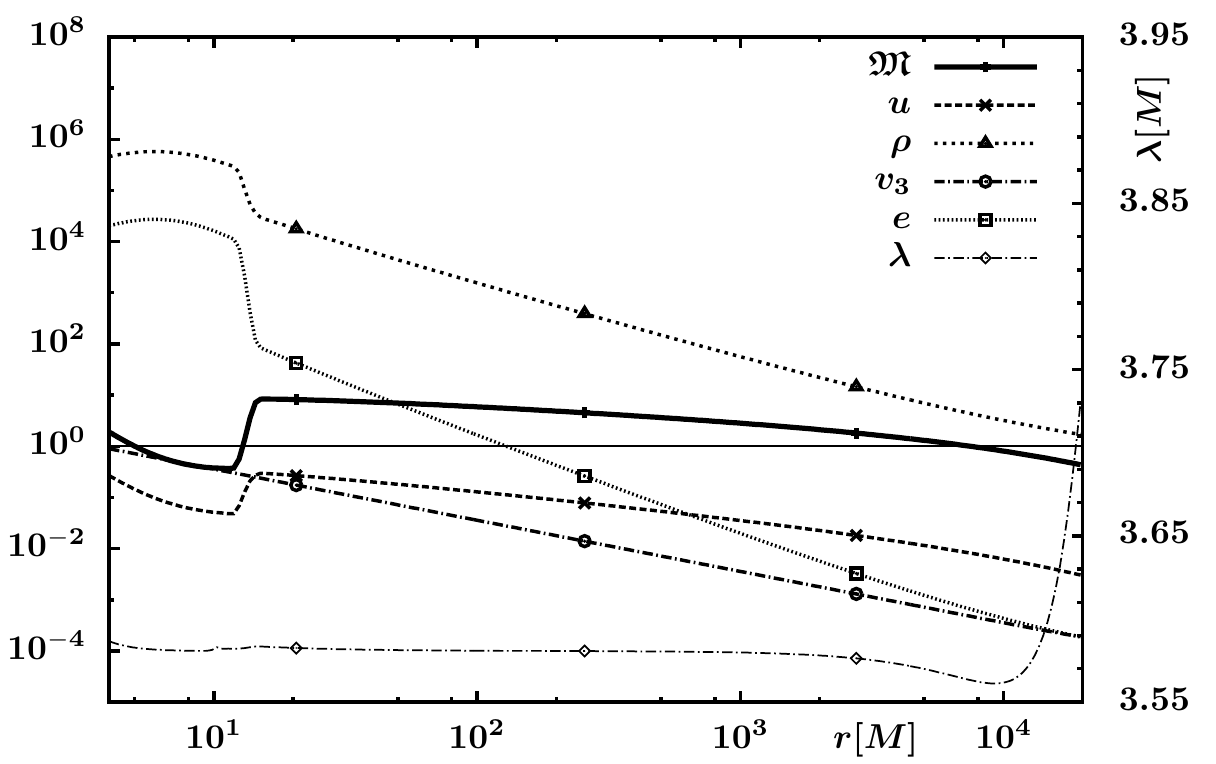}
\includegraphics[width=0.495\textwidth]{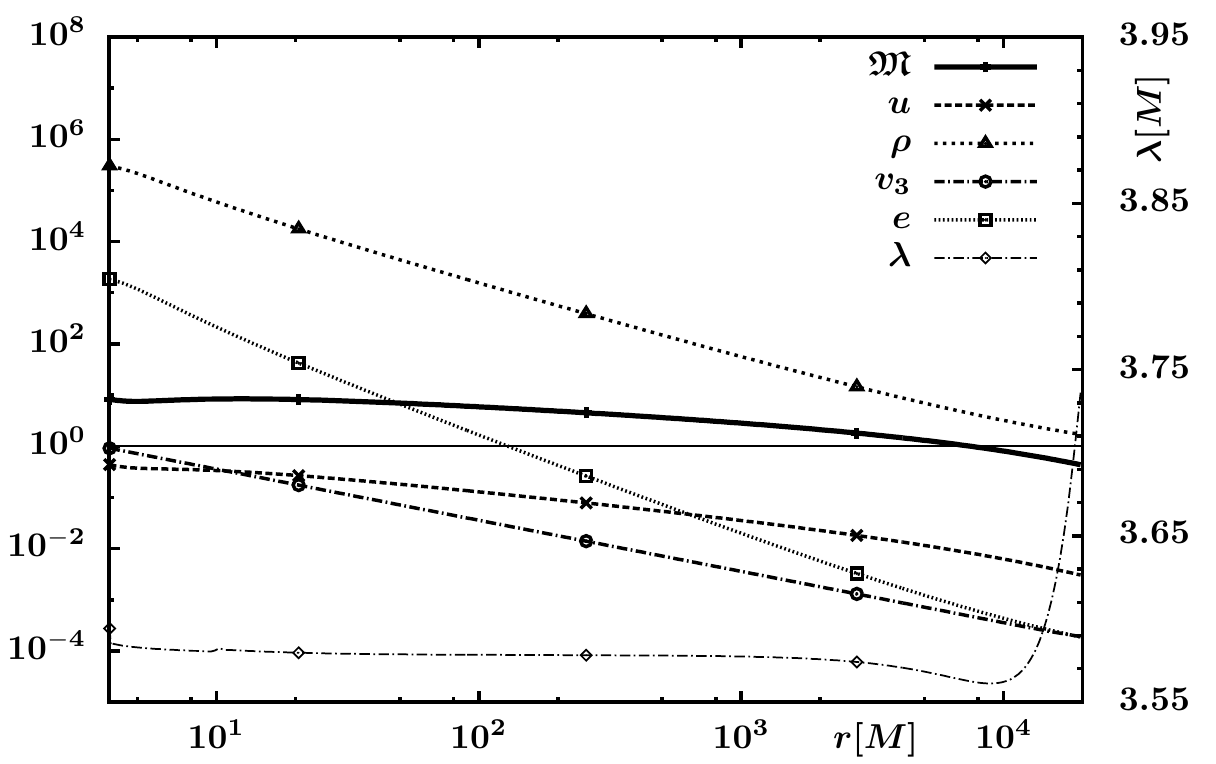}

\includegraphics[width=0.495\textwidth]{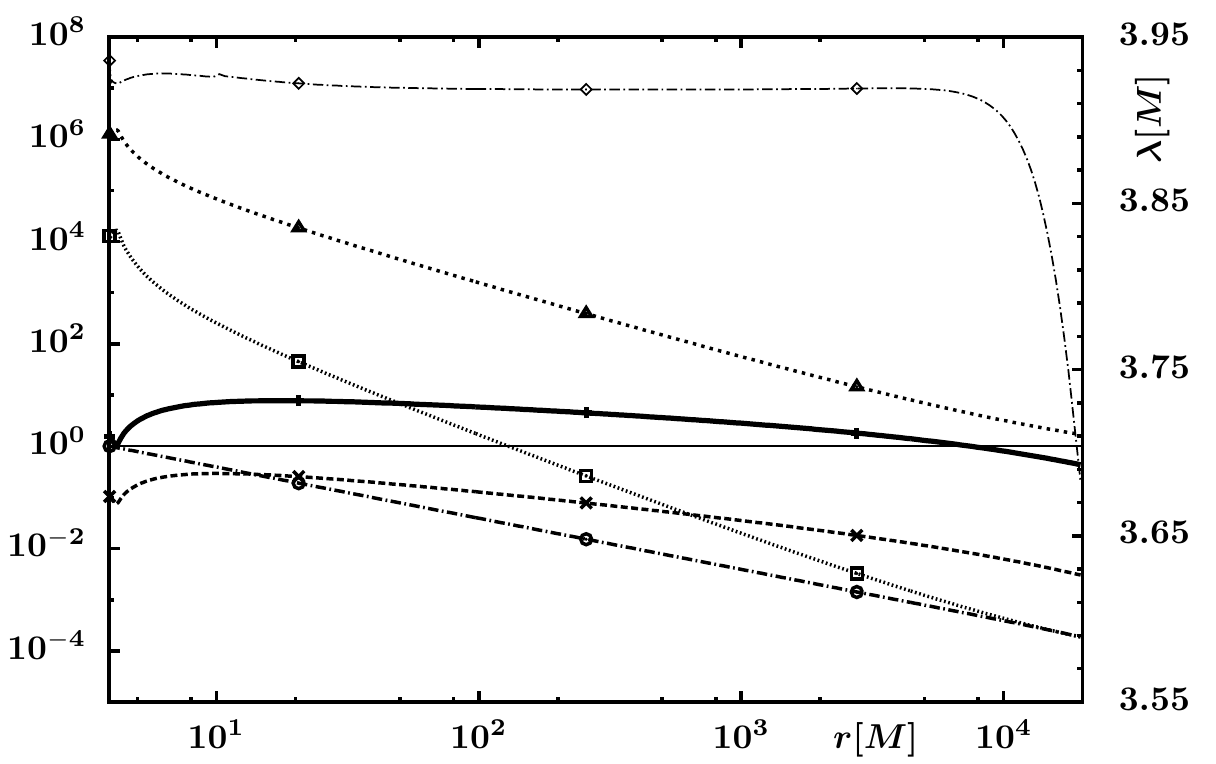}
\includegraphics[width=0.495\textwidth]{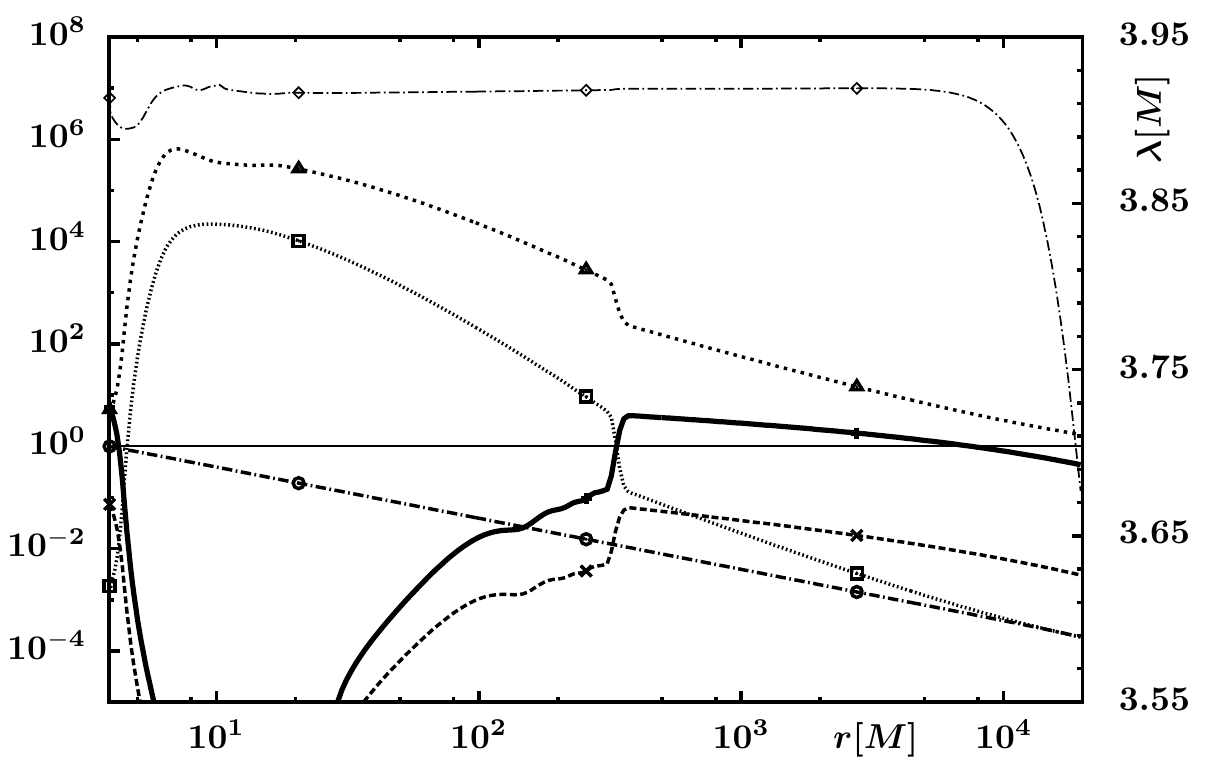}

\includegraphics[width=0.495\textwidth]{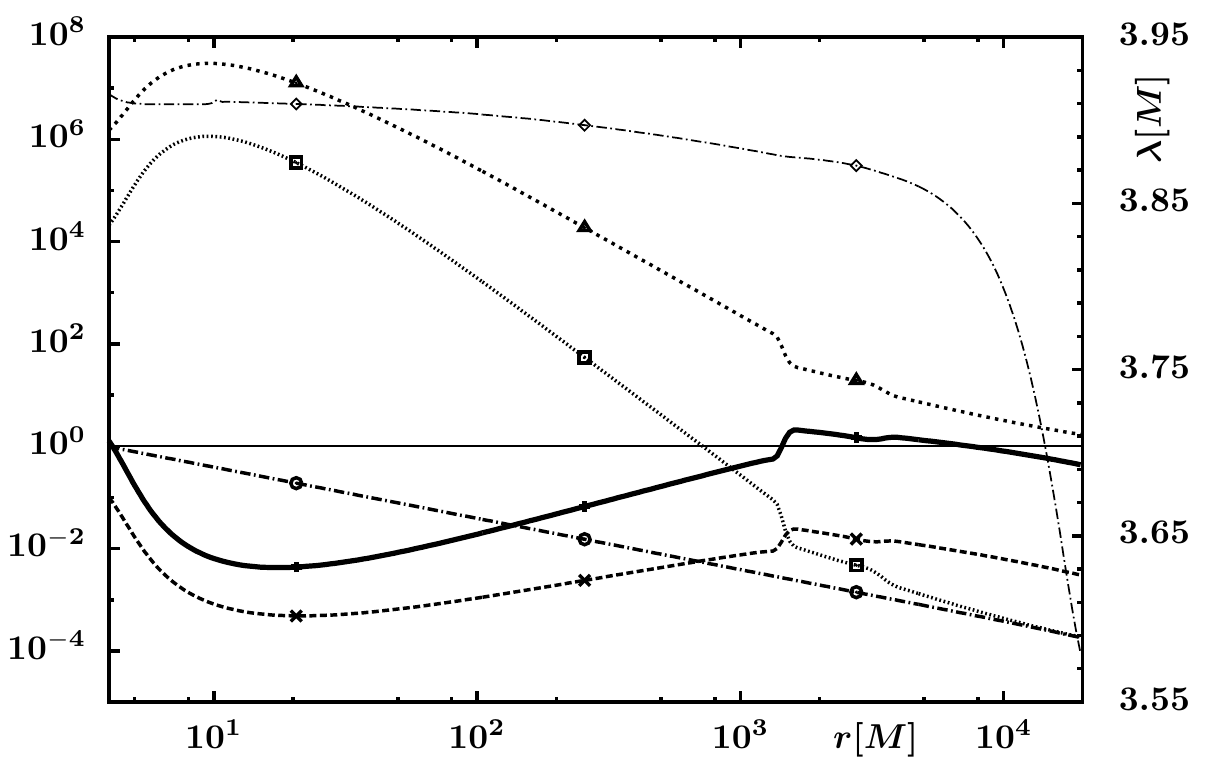}
\includegraphics[width=0.495\textwidth]{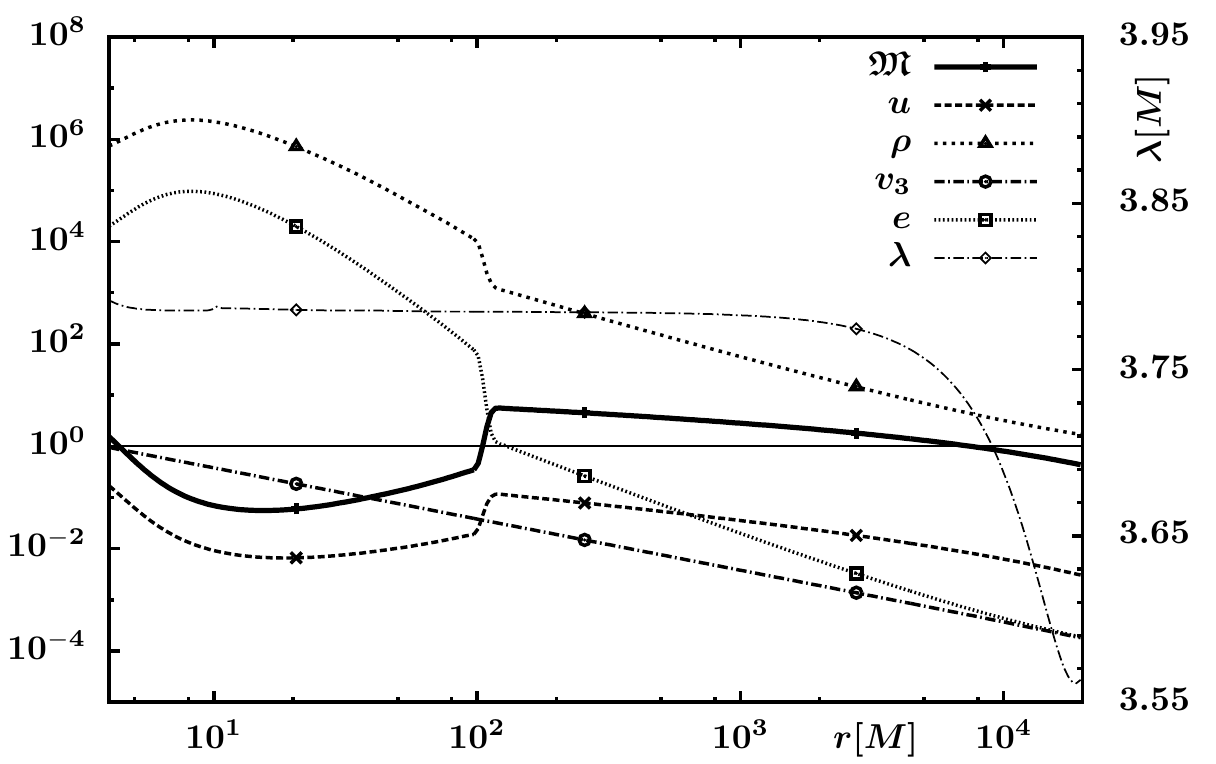}

\caption
{Normalized accretion rate through the inner boundary (top left) and the zoom of the smaller peak (top right). Snapshot of the accretion flow profiles at times $t_1=2.01\cdot 10^7M$ (2. row left), $t_2=2.015 \cdot 10^7M$ (2. row right), $t_3=2.57 \cdot 10^7M$ (3. row left), $t_4=2.575 \cdot 10^7M$ (3. row right), $t_5=2.695 \cdot 10^7M$ (4. row left) and $t_6=2.795 \cdot 10^7M$ (4. row right). The meaning of the plots is the same as in Fig. \ref{fig_ha_hn}, moreover the angular momentum $\lambda(r)$ is also shown. The parameters are $\gamma=4/3$, $\epsilon=0.0001$, $\lambda_0=3.74M$, $\Delta\lambda=0.18M$, $T_v=1\cdot 10^7M$.
In the zoomed picture (top right) also two similar peaks with $\epsilon=0.0000033$ are shown with the right $y$-axis ($\dot{M}^2: \gamma=1.5$  and $\dot{M}^3: \gamma=1.6$), the quantities are denoted by the superscript. The time axis $\tilde{t}=t-t_0$ is shifted by $t_0^1=10168175M$, $t_0^2=17778500M$ and $t_0^3=8951542M$. (The animation is accessible  in the online version.)
\label{fig_qp}
}
\end{figure*}


\section{Concluding remarks} \label{conclusion}

In this paper we derive the accretion flow equation and shock condition which hold for quasi-spherical slightly rotating flow. Using these equations we find numerically the accretion solution with standing Rankine-Hugoniot shock and we input it as the initial condition for simulation of the flow in the hydrodynamical code ZEUS. The relativistic effects of the massive compact objects in the center are taken into account by means of the pseudo-Newtonian Paczynski-Wiita potential. This {\color{black} or similar} configuration {\color{black} with low angular momentum} could be relevant in low luminous sources, such as non-active galactic nuclei or X-ray binaries in their quiescent state, or the spherical coronal inflow, responsible for hard X-ray emission of the black hole binaries in their hard and soft 
states. 

At long distances our steady solution converges to {\color{black} that of spherical accretion flow} \citep{1952MNRAS.112..195B} with $a_\infty$ given by (\ref{a_infty}) and with the Bondi accretion rate. 
We follow the dynamical evolution of the flow for long integration time and we observe oscillation of the shock position for certain subset of parameters allowing the shock solution. {\color{black} Because no cooling is considered in our model, the underlying physical reason for the oscillations seen in the numerical simulations is yet unclear.  We checked that this effect is not caused by the discretisation or by different initial conditions and that the oscillations develop for the same set of parameters. However, the value of frequency and amplitude can in principle depend slightly on the discretisation, because the range of radii reached by the oscillating shock front is typically covered by relatively low number of grid zones ($\sim$ 10 -- 20), hence the oscillations are not resolved in detail. 

The possible numerical origin of these oscillations could be also connected with the numerical viscosity, which is present in the ZEUS algorithm. For the presented results we used {\tt qcon=2.0} and {\tt qlin=0.0}, which are the ZEUS parameters describing the quadratic and linear viscosity, respectively (typically the shock is smeared into {\tt qcon+2} zones by the algorithm). The presence of the numerical viscosity is needed to ensure the correct production of entropy  and the corresponding jump in entropy accretion rate at the shock front in the system of formally isentropic equations describing the motion of the fluid \citep{2010ApJS..187..119C}.  We carried out several test simulations with different reasonable values of these parameters ({\tt qcon} $\in [1,4]$, {\tt qlin} $\in [0,0.5]$, which doubles the  range mentioned in \cite{2010ApJS..187..119C}). We found changes of frequency approx. within 5\% with the changing artificial viscosity parameters. Because the assumption of purely inviscid accretion is unphysical,  the role of both the physical and numerical viscosity on emergence of the oscillations and its influence on their frequency and amplitude need to be investigated further.

The oscillations of the shock front}  also lead to the variation of accretion rate on the center. The amplitude of these oscillation and accordingly the amplification factor of the accretion rate $F=\dot{M}_{\rm max}/\dot{M}_{\rm min}$ is between one and ten. The frequency of the oscillations depends on the specific angular momentum of the flow and forms two well defined branches. However, the energy dependence of the shock position and the frequency of oscillation is relatively weak.

We also followed the evolution of the flow when the value of specific angular momentum of the material incoming through the outer boundary is slowly periodically changing. In this case for some values of parameters quite high and sharp peak in the accretion rate is observed with the ratio between the peak half-width and the variation period in the order $10^{-3}$. The shape of the peak agrees with the observed flares of Sgr ~A* with slower rise and sharp decline, but the time scale of the peak for the used parameters is longer than the  observed ones. 

We also observe much sharper peaks which correspond to the situation when the leading parameter is going from the shock solution interval to the region where only the solution passing through the outer sonic point exists. The shorter time scale is connected with the lowest possible position of the shock before this transition ${r_s}_{\rm min} \approx (10 \div 40) M$. The value of ${r_s}_{\rm min}$ is mainly given by the polytropic constant $\gamma$. The width of these peaks is on the order of hundreds $M$ ($50M\div250M)$ which agrees well with the time scale of the observed flares from Sgr~A* ($80\div270M$), {\color{black} provided that the angular momentum of Sgr A* accreting medium is low.}

For variation of the leading parameter wider than the shock existence interval we observe the hysteresis effect proposed by \cite{2012NewA...17..254D}. During the hysteresis loop both types of peaks together with slow modulation of the accretion rate are present. {\color{black} This scenario also includes repeating creation and disappearance of the shock in the flow.}

Because the time of these processes scales linearly with the mass of the accreting object, we can separately treat the case of stellar black holes and super-massive black holes in quiescent Galaxies. 
In case of supermassive black hole, the time scale is quite long. For Sgr~A*, our closest supermassive black hole, $1M\approx21.5s$, hence the longer peaks duration, which we obtained, corresponds to 2 to 50 days and can be longer depending on $T_v$ and the short peak duration is in $ks$ depending on adiabatic index $\gamma$ but independent on $T_v$. For supermassive black holes the short peak is observable and {\color{black} in agreement} with the duration of flares of Sgr~A*.

On the other hand for stellar black holes with typical mass $M=10M_\odot$, the unit of time $1M \approx 5\cdot10^{-5}s$. The typical frequencies in Hz of the shock position oscillations are depicted on the top axis of the spectra and correspond quite well with the QPOs measured in some sources, e.g. 1915+105 or GX 339-4. Eventhough the former particular source is not a quiescent source and the Keplerian accretion disc is probably present, the hot corona flow can be quasi-spherical. {\color{black} If it contains only small amount of angular momentum, similar shock oscillations can be generated in the appropriate low angular momentum region}.

The results from Paragraph \ref{loop} cannot be directly applied to {\color{black} observations of} stellar black holes because the timescale of the obtained peaks in the range $(1\div20)\cdot 10^4M \approx (0.5\div10)s$ for the longer peaks and $100M \approx 0.005s$ for the shorter peaks. However the value of the variational period $T_v$ was chosen in such way to obtain results in reasonable computer time, but actually as we mentioned earlier in case of stellar black holes the orbital period of the companion could be much longer \citep{2014arXiv1406.7262G}.

In this context we can mention the observation of flares in the source GX 339-4 by \cite{2012A&A...542A..56N}. This source is a black hole candidate with the estimated mass $M=5.8\pm0.5M_\odot$ and it undergoes several outburst with the time scale of two to three years. The authors reported about detection of QPOs with increasing frequency from 0.102 Hz to 5.692 Hz within 26 days. After that period sporadic or none detection of QPOs were made. Than during the decline the QPOs were detected again with decreasing frequency from 6.42Hz to 1.149Hz. The authors linked this behaviour with the propagating oscillatory shock model and presented a variation of the shock position with time. {\color{black} Within our very simplified model} we can understand the change of QPO frequency in view of changing angular momentum of the flow with very large variational period $T_v$. {\color{black} Even though we did not consider many physical effects, like viscosity of the flow, a more complicated geometrical structure with possible presence of the Keplerian accretion disc or the presence of magnetic fields}, the observed range of frequencies is in very good agreement with our results. {\color{black} Particularly,} for $\epsilon=0.0001$, $\gamma=4/3$ and $M=5.8M_\odot$ the lowest obtained frequency in our simulations is 0.084 Hz for $\lambda=3.925M$, for $\lambda=3.905M$ the frequency of oscillations is 0.105 Hz and the highest obtained frequency is 5.6 Hz for $\lambda=3.805M$. {\color{black} Although our simple model cannot describe accurately the real ongoing accretion, this agreement between the observed and obtained frequencies could  indicate that similar behaviour of the low angular momentum layer or quasi-spherical corona is present in the astrophysical systems. }

{\color{black} In the future we plan to incorporate more of these aspects in our model. Our further studies will be based on higher dimensional computations to take into account the geometrical structure of the accretion flow. Also, the non-zero viscosity of the real gas influences the properties of the dynamical system. The viscous effects on the flow were studied by \cite{1990MNRAS.243..610C} who showed that the presence of viscous dissipation redistributes the angular momentum and may remove the multiple shock locations, as the geometrically thin disk accretion is possible. Nevertheless,
 the overall phase portrait remains similar and shock solutions are possible, provided the $\alpha$ viscosity is smaller than some critical value. The values of viscous  parameter $\alpha$ used in their study are comparable with viscosities of convective or magnetic origin, whereas the radiative and molecular viscosities are orders of magnitude lower, hence these types of  viscous interaction cannot change the phase portrait of the flow. 
Nevertheless, it is important to note that overall 
$\alpha$ is an average value over the disk
height, and therefore it has no clear meaning in the case of a 
geometrically thick flow. 
The exact values that can be verified observationally, are reliable only in the case 
of a geometrically thin accretion disk models (see \cite{2007MNRAS.376.1740K}). 
}

\section*{Acknowledgements}

We thank Bozena Czerny, Tapas Das and Marat Gilfanov for helpful discussions.
This work was supported in part by 
grant DEC-2012/05/E/ST9/03914 from the Polish National Science Center.

\bibliography{Sukova}

\begin{thebibliography}{}

\bibitem[\protect\citeauthoryear{{Abramowicz} \& {Chakrabarti}}{{Abramowicz} \&
  {Chakrabarti}}{1990}]{1990ApJ...350..281A}
{Abramowicz} M.~A.,  {Chakrabarti} S.~K.,  1990, ApJ, 350, 281

\bibitem[\protect\citeauthoryear{{Abramowicz} \& {Zurek}}{{Abramowicz} \&
  {Zurek}}{1981}]{1981ApJ...246..314A}
{Abramowicz} M.~A.,  {Zurek} W.~H.,  1981, ApJ, 246, 314

\bibitem[\protect\citeauthoryear{{Bondi}}{{Bondi}}{1952}]{1952MNRAS.112..195B}
{Bondi} H.,  1952, MNRAS, 112, 195

\bibitem[\protect\citeauthoryear{{Chakrabarti} \& {Titarchuk}}{{Chakrabarti} \&
  {Titarchuk}}{1995}]{1995ApJ...455..623C}
{Chakrabarti} S.,  {Titarchuk} L.~G.,  1995, ApJ, 455, 623

\bibitem[\protect\citeauthoryear{{Chakrabarti}}{{Chakrabarti}}{1989}]{1989ApJ...347..365C}
{Chakrabarti} S.~K.,  1989, ApJ, 347, 365

\bibitem[\protect\citeauthoryear{{Chakrabarti}}{{Chakrabarti}}{1990}]{1990MNRAS.243..610C}
{Chakrabarti} S.~K.,  1990, MNRAS, 243, 610

\bibitem[\protect\citeauthoryear{{Chakrabarti} \& {Manickam}}{{Chakrabarti} \&
  {Manickam}}{2000}]{2000ApJ...531L..41C}
{Chakrabarti} S.~K.,  {Manickam} S.~G.,  2000, ApJ Lett., 531, L41

\bibitem[\protect\citeauthoryear{{Chaudhury}, {Ray} \& {Das}}{{Chaudhury}
  et~al.}{2006}]{2006MNRAS.373..146C}
{Chaudhury} S.,  {Ray} A.~K.,    {Das} T.~K.,  2006, MNRAS, 373, 146

\bibitem[\protect\citeauthoryear{{Clarke}}{{Clarke}}{2010}]{2010ApJS..187..119C}
{Clarke} D.~A.,  2010, ApJS, 187, 119

\bibitem[\protect\citeauthoryear{{Cuadra}, {Nayakshin} \& {Martins}}{{Cuadra}
  et~al.}{2008}]{2008MNRAS.383..458C}
{Cuadra} J.,  {Nayakshin} S.,    {Martins} F.,  2008, MNRAS, 383, 458

\bibitem[\protect\citeauthoryear{{Czerny} \& {Mo{\'s}cibrodzka}}{{Czerny} \&
  {Mo{\'s}cibrodzka}}{2008}]{2008JPhCS.131a2001C}
{Czerny} B.,  {Mo{\'s}cibrodzka} M.,  2008, Journal of Physics Conference
  Series, 131, 012001

\bibitem[\protect\citeauthoryear{{Das}}{{Das}}{2002}]{2002ApJ...577..880D}
{Das} T.~K.,  2002, ApJ, 577, 880

\bibitem[\protect\citeauthoryear{{Das}}{{Das}}{2003}]{2003ApJ...588L..89D}
{Das} T.~K.,  2003, ApJ Lett., 588, L89

\bibitem[\protect\citeauthoryear{{Das} \& {Czerny}}{{Das} \&
  {Czerny}}{2012}]{2012NewA...17..254D}
{Das} T.~K.,  {Czerny} B.,  2012, NA, 17, 254

\bibitem[\protect\citeauthoryear{{Das}, {Rao} \& {Vadawale}}{{Das}
  et~al.}{2003}]{2003MNRAS.343..443D}
{Das} T.~K.,  {Rao} A.~R.,    {Vadawale} S.~V.,  2003, MNRAS, 343, 443

\bibitem[\protect\citeauthoryear{{Di Matteo}, {Allen}, {Fabian}, {Wilson} \&
  {Young}}{{Di Matteo} et~al.}{2003}]{2003ApJ...582..133D}
{Di Matteo} T.,  {Allen} S.~W.,  {Fabian} A.~C.,  {Wilson} A.~S.,    {Young}
  A.~J.,  2003, ApJ, 582, 133

\bibitem[\protect\citeauthoryear{{Genzel}}{{Genzel}}{2003}]{2003HEAD....7.0301G}
{Genzel} R.,  2003, in AAS/High Energy Astrophysics Division \#7 Vol.~35 of
  Bulletin of the American Astronomical Society, {The Galactic Center Black
  Hole}.
p.~606

\bibitem[\protect\citeauthoryear{{Ghosh} \& {Chakrabarti}}{{Ghosh} \&
  {Chakrabarti}}{2014}]{2014arXiv1406.7262G}
{Ghosh} A.,  {Chakrabarti} S.~K.,  2014, ArXiv e-prints

\bibitem[\protect\citeauthoryear{{Gilfanov}}{{Gilfanov}}{2010}]{2010LNP...794...17G}
{Gilfanov} M.,  2010, in {Belloni} T.,  ed., Lecture Notes in Physics, Berlin
  Springer Verlag Vol.~794 of Lecture Notes in Physics, Berlin Springer Verlag,
  {X-Ray Emission from Black-Hole Binaries}.
p.~17

\bibitem[\protect\citeauthoryear{{Hayes} \& {Norman}}{{Hayes} \&
  {Norman}}{2003}]{2003ApJS..147..197H}
{Hayes} J.~C.,  {Norman} M.~L.,  2003, ApJS, 147, 197

\bibitem[\protect\citeauthoryear{{Janiuk}, {Proga} \& {Kurosawa}}{{Janiuk}
  et~al.}{2008}]{2008ApJ...681...58J}
{Janiuk} A.,  {Proga} D.,    {Kurosawa} R.,  2008, ApJ, 681, 58

\bibitem[\protect\citeauthoryear{{Janiuk}, {Sznajder}, {Mo{\'s}cibrodzka} \&
  {Proga}}{{Janiuk} et~al.}{2009}]{2009ApJ...705.1503J}
{Janiuk} A.,  {Sznajder} M.,  {Mo{\'s}cibrodzka} M.,    {Proga} D.,  2009, ApJ,
  705, 1503

\bibitem[\protect\citeauthoryear{{King}, {Pringle} \& {Livio}}{{King}
  et~al.}{2007}]{2007MNRAS.376.1740K}
{King} A.~R.,  {Pringle} J.~E.,    {Livio} M.,  2007, MNRAS, 376, 1740

\bibitem[\protect\citeauthoryear{{Loeb}}{{Loeb}}{2004}]{2004MNRAS.350..725L}
{Loeb} A.,  2004, MNRAS, 350, 725

\bibitem[\protect\citeauthoryear{Meyer, Ghez, Schödel, Yelda, Boehle, Lu, Do,
  Morris, Becklin \& Matthews}{Meyer et~al.}{2012}]{Meyer05102012}
Meyer L.,  Ghez A.~M.,  Schödel R.,  Yelda S.,  Boehle A.,  Lu J.~R.,  Do T.,
  Morris M.~R.,  Becklin E.~E.,    Matthews K.,  2012, Science, 338, 84

\bibitem[\protect\citeauthoryear{{Meyer-Hofmeister} \&
  {Meyer}}{{Meyer-Hofmeister} \& {Meyer}}{2014}]{2014A&A...562A.142M}
{Meyer-Hofmeister} E.,  {Meyer} F.,  2014, A\&A, 562, A142

\bibitem[\protect\citeauthoryear{{Mo{\'s}cibrodzka}, {Das} \&
  {Czerny}}{{Mo{\'s}cibrodzka} et~al.}{2006}]{2006MNRAS.370..219M}
{Mo{\'s}cibrodzka} M.,  {Das} T.~K.,    {Czerny} B.,  2006, MNRAS, 370, 219

\bibitem[\protect\citeauthoryear{{Nag}, {Acharya}, {Ray} \& {Das}}{{Nag}
  et~al.}{2012}]{2012NewA...17..285N}
{Nag} S.,  {Acharya} S.,  {Ray} A.~K.,    {Das} T.~K.,  2012, NA, 17, 285

\bibitem[\protect\citeauthoryear{{Nandi}, {Debnath}, {Mandal} \&
  {Chakrabarti}}{{Nandi} et~al.}{2012}]{2012A&A...542A..56N}
{Nandi} A.,  {Debnath} D.,  {Mandal} S.,    {Chakrabarti} S.~K.,  2012, A\&A,
  542, A56

\bibitem[\protect\citeauthoryear{{Narayan} \& {Fabian}}{{Narayan} \&
  {Fabian}}{2011}]{2011MNRAS.415.3721N}
{Narayan} R.,  {Fabian} A.~C.,  2011, MNRAS, 415, 3721

\bibitem[\protect\citeauthoryear{{Narayan} \& {Yi}}{{Narayan} \&
  {Yi}}{1994}]{1994ApJ...428L..13N}
{Narayan} R.,  {Yi} I.,  1994, ApJ, 428, L13

\bibitem[\protect\citeauthoryear{{Nowak}, {Neilsen}, {Markoff}, {Baganoff},
  {Porquet}, {Grosso}, {Levin}, {Houck}, {Eckart}, {Falcke}, {Ji}, {Miller} \&
  {Wang}}{{Nowak} et~al.}{2012}]{2012ApJ...759...95N}
{Nowak} M.~A.,  {Neilsen} J.,  {Markoff} S.~B.,  {Baganoff} F.~K.,  {Porquet}
  D.,  {Grosso} N.,  {Levin} Y.,  {Houck} J.,  {Eckart} A.,  {Falcke} H.,  {Ji}
  L.,  {Miller} J.~M.,    {Wang} Q.~D.,  2012, ApJ, 759, 95

\bibitem[\protect\citeauthoryear{{Okuda} \& {Molteni}}{{Okuda} \&
  {Molteni}}{2012}]{2012MNRAS.425.2413O}
{Okuda} T.,  {Molteni} D.,  2012, MNRAS, 425, 2413

\bibitem[\protect\citeauthoryear{{Proga} \& {Begelman}}{{Proga} \&
  {Begelman}}{2003}]{2003ApJ...582...69P}
{Proga} D.,  {Begelman} M.~C.,  2003, ApJ, 582, 69

\bibitem[\protect\citeauthoryear{{Proga} \& {Zhang}}{{Proga} \&
  {Zhang}}{2006}]{2006MNRAS.370L..61P}
{Proga} D.,  {Zhang} B.,  2006, MNRAS, 370, L61

\bibitem[\protect\citeauthoryear{{Shakura} \& {Sunyaev}}{{Shakura} \&
  {Sunyaev}}{1973}]{1973A&A....24..337S}
{Shakura} N.~I.,  {Sunyaev} R.~A.,  1973, A\&A, 24, 337

\bibitem[\protect\citeauthoryear{{Stone} \& {Norman}}{{Stone} \&
  {Norman}}{1992}]{1992ApJS...80..753S}
{Stone} J.~M.,  {Norman} M.~L.,  1992, ApJS, 80, 753

\bibitem[\protect\citeauthoryear{{Yuan} \& {Narayan}}{{Yuan} \&
  {Narayan}}{2014}]{2014ARA&A..52..529Y}
{Yuan} F.,  {Narayan} R.,  2014, ARA\&A, 52, 529

\end{thebibliography}

\end{document}